\documentclass{aa}  

\usepackage{graphicx}
\usepackage{txfonts}
\usepackage{natbib}
\usepackage{color}
\newcommand{\nc}{\newcommand}
\def\cplus{C$^{+}$}
\def\thcplus{$^{13}$C$^{+}$}
\nc{\Msun}{\ensuremath{\mathrm{M}_\odot}}
\nc{\Rsun}{\ensuremath{\mathrm{R}_\odot}}
\nc{\cmcub}{\mbox{cm$^{-3}$}}
\newcommand{\CII}{[C {\sc ii}]}
\newcommand{\thCII}{[$^{13}$C {\sc ii}]}

\newcommand{\HI}{H {\sc i}}
\nc\micron{\mbox{$\mu$m}}
\nc{\thCO}{$^{13}$CO}
\nc{\hcoplus}{HCO$^+$}
\newcommand{\OI}{[O {\sc i}]}
\newcommand{\vlsr}{$\upsilon_{\rm LSR}$}
\nc{\cmsq}{\mbox{cm$^{-2}$}}
\nc{\kms}{\mbox{km~s$^{-1}$}}
\nc{\lsun}{\ensuremath{\mathrm{L}_\odot}}
\newcommand{\HII}{H {\sc ii}}
\nc{\CeiO}{C$^{18}$O}
\nc{\tex}{T$_{\rm ex}$}
\nc{\tkin}{T$_{\rm kin}$}
\nc{\rsun}{\ensuremath{\mathrm{R}_\odot}}

\newcommand\phn{\phantom{0}}%

\def\ptsec{$''\mskip-7.6mu.\,$}

\begin{document}

   \title{The distribution of ionized, atomic and PDR gas
   around S\,1\\ in $\rho$\,Ophiuchus}

   \author{B. Mookerjea \inst{1} 
       \and  G. Sandell \inst{2}
      \and    V. S. Veena \inst{3} 
   \and       R. G\"usten \inst{4}
   \and D. Riquelme \inst{4}
   \and H. Wiesemeyer \inst{4}
   \and       F. Wyrowski \inst{4}
   \and       M. Mertens \inst{3}
          }

   \institute{Tata Institute of Fundamental Research, Homi Bhabha Road,
Mumbai 400005, India
              \email{bhaswati@tifr.res.in}
         \and
         Institute for Astronomy, University of Hawaii,  640 N. Aohoku Place, Hilo, HI 96720, USA
	 \and
	 I. Physikalisches Institut, Universit\"at zu K\"oln, Z\"ulpicher Str. 77, D-50937 K\"oln, Germany
         \and
Max Planck Institut f\"ur Radioastronomie, Auf dem H\"ugel 69, D-53121 Bonn, Germany 
             }

   \date{latest revision \today}

\abstract {The early B star S\,1 in the $\rho$ Ophiuchus cloud excites
an \HII\ region and illuminates a large egg-shaped photodissociation
(PDR) cavity. The PDR is restricted to the west and south-west by the
dense molecular $\rho$\,Oph A ridge, expanding more freely into the
diffuse low density cloud to the north-east.  We analyze new SOFIA
GREAT, GMRT and APEX data together with archival data from
Herschel/PACS, JCMT/HARPS to study the properties of the
photo-irradiated ionized and neutral gas in this region. The tracers
include \CII\ at 158\,\micron, \OI\ at 63 and 145\,\micron, $J$=6--5
transitions of CO and \thCO, \hcoplus (4--3), radio continuum at 610 and
1420\,MHz and \HI\ at 21\,cm. The PDR emission is strongly red-shifted
to the south-east of the nebula, and primarily blue-shifted on the north
western side.  The \CII\ and and \OI63\ spectra are strongly
self-absorbed  over most of the PDR. By using the optically thin
counterparts, \thCII\ and \OI145\ respectively, we conclude that the
self-absorption is dominated by the warm ($>80$\,K) foreground PDR gas
and not  by the surrounding cold molecular cloud. We estimate the column
densities of \cplus\ and O$^0$ of the PDR to be $\sim 3\times 10^{18}$
and $\sim 2\times 10^{19}$\,\cmsq, respectively. Comparison of stellar
far-ultraviolet flux and reprocessed infrared radiation suggest enhanced
clumpiness of the gas to the north-west. Analysis of the emission from
the PDR gas suggests the presence of at least three density components
consisting of high density (10$^6$\,\cmcub) clumps, medium density
(10$^4$\,\cmcub) and diffuse (10$^3$\,\cmcub) interclump medium.  The
medium density component primarily contributes to the thermal pressure
of the PDR gas which is in pressure equilibrium with the molecular cloud
to the west. Emission velocities in the region suggest that the PDR is
tilted and somewhat warped with the south-eastern side of the cavity
being denser on the front and the north-western side being denser on the
rear.  }


\maketitle
%

\section{Introduction}

Photon-dominated Regions (PDRs)  are regions where far-ultraviolet (FUV)
(6\,eV $<$ h$\nu <$ 13.6 eV) radiation from young massive stars dominate
the physics and the chemistry of the interstellar medium
\citep{Tielens85}.  The PDRs play an important role in reprocessing much
of the energy from stars and re-emitting this energy in the
infrared-millimeter regime. Most of the mass of the gas and dust in the
Galaxy resides in PDRs \citep{Hollenbach1999}. In the far infrared the
most important cooling lines are the fine structure lines of \CII\ at
158\,\micron, and \OI\ at 63 \& 145\,\micron\ and to a lesser extent
high-$J$ CO lines, while PAH emission and H$_2$ lines dominate in the
near- and mid-IR. Of these tracers \CII\ being the most abundant and
easily excited is the most ubiquitous. Owing to an ionization potential
of 11.26\,eV for Carbon, understanding the phase of gas from which \CII\
arises requires comparison with bona-fide tracers of ionized, atomic and
molecular gas. 

\citet{mookerjea2018} recently published an observational study of the
\CII\ emission from a region around the B4V star S\,1 located in the
$\rho$ Ophiuchus dark cloud \citep[at a distance of
137.3$\pm$1.2\,pc]{Ortiz17}. The S\,1 PDR is located on the eastern edge
of the westernmost core $\rho$\,Oph\,A \citep{loren90}.  The
$\rho$\,Oph\,A core  has a filamentary structure with at least nine pre-
and protostellar cores \citep{wilson99,difrancesco04}, including the
prototypical Class 0 source VLA\,1623.  Most of the cores are starless,
although two of the cores may have embedded protostars
\citep{friesen18,kawabe18}.  \citet{mookerjea2018} found that the  \CII\
emission is dominated by the strong emission from  the nebula
surrounding S\,1 that appears to expand into the dense Oph\,A molecular
cloud to the west and south of S\,1.  The \CII\ emission is distributed
similar to the other PDR tracers such as the 8\,\micron\ continuum
tracing emission from PAHs and the velocity-integrated emission of \OI\
at 145 and 63\,\micron\ measured by \citet{larsson2017}.  A comparison
of \CII\ with the $J$=3--2 emission of CO and $^{13}$CO shows very
little similarity, although the highly compressed parts of the PDR shell
traced by \CII\ show up in \CeiO(3--2) as well as HCO$^+$(4--3).
\citet{mookerjea2018} also detected \CII\ to be strongly self-absorbed
over an extended region in the S\,1 PDR and interpreted it as a cold
foreground cloud being absorbed against the warm background gas.
Analysing velocity-unresolved Herschel/PACS data, \citet{larsson2017}
deduced that this cold foreground cloud absorbs most of the
$^3$P$_{1}$--$^3$P$_{2}$ \OI\ 63\,\micron\ radiation but leaves the
higher level $^3$P$_{0}$--$^3$P$_{1}$ \OI\ 145\,\micron\ line
unaffected.

In this paper we present newly observed maps of radio continuum, \HI\ at
21\,cm, \CII\ at 158\,\micron, \OI\ at 63 and 145\,\micron\ and $J$=6--5
transitions of CO and \thCO\, and use them to study the morphology and
physical properties of the PDR around the star S\,1.

\section{Observations \& Data Reduction}

\subsection{SOFIA}

The S\,1 PDR was observed on two occasions on June 14, 2018
and June 5, 2019 with upGREAT\footnote{The German REceiver for
Astronomy at Terahertz frequencies (upGREAT) is a development
by the MPI f\"ur Radioastronomie and the KOSMA/Universit\"at
zu K\"oln, in cooperation with the DLR Institut f\"ur Optische
Sensorsysteme.} \citep{Risacher2018} on flights leaving from
Christchurch, New Zealand. All observations were done in
Consortium time (Project 83\_0614). The bright PDR was mapped
simultaneously in both oxygen fine-structure lines: the HFA
array was tuned to \OI\ 63 $\mu$m (f = 4744.77749 GHz), while
the LFA-H polarization sub-array  was tuned to \OI\
145\,\micron\ (f = 2060.06886 GHz). The mapped field,
indicated in Fig.4, was sampled at 3\arcsec\ spacing, with 0.4
sec integration time per dump. In order to record the
extended, lower-level PDR emission, in 2019 a wider field of
294\arcsec\ $\times$ 294\arcsec\, centered on S\,1, was added
with the LFA tuned to \CII\ 158\,\micron\, sampling every
6\arcsec\ at a scan rate of 0.4 sec per resolution element.
All mapping was carried-out under dry atmospheric conditions
at 42,000--43,000 ft flight altitude in total power on-the-fly
mode, with the reference position at -120\arcsec, +300\arcsec\
relative to S\,1.  The off position was clean for \OI, but
there was still \CII\ emission in the off position.  We
therefore took a longer single pointed observation toward this
off against a far off position (offset at
833\arcsec,-167\arcsec), which allowed us to the correct the
\CII\ map for the contamination in the (near) off position.
The spectrometer setting during the \CII\ observation also
covered the strongest hyperfine transition (2--1;
1900.4661\,Hz) of the \thCII.

The observations were reduced and calibrated by the GREAT team.
The GREAT team also provided beam sizes (14\farcs1 for \CII,
13\arcsec\ for \OI145 and 6\farcs3 for \OI63) and beam
efficiencies derived from planet observations.  The data were
corrected for atmospheric extinction and calibrated in T$_{mb}$.
In June 2018 the telluric 63 $\mu$m \OI\ line was at  V$_{\rm
lsr}$ = 1.6 \kms, essentially making the 63\,micron\ data
unusable. In June 2019 the S\,1 PDR  was observed earlier in the
month shifting the telluric 63 $\mu$m \OI\ to 7.5 \kms, well
away from the emission from the PDR except for some of the
red-shifted \OI\ spectra in the southern part of the PDR
cavity.  Comparison with the 2018 data which were clean at these
velocities, show that very few spectra in the 2019 data are
affected. Further processing of the data (conversion to
main-beam brightness temperature, with beam efficiencies of
0.58, and averaging with 1/$\sigma^2$ rms weighting) was made
with the CLASS\footnote{CLASS is part of the GILDAS software
package, see http://www.iram.fr/IRAMFR/GILDAS} software. The
final maps of \CII\ and \OI\ 63 and 145\,\micron\ presented here
are centered at 16:26:34.175 -24:23:28.3 (J2000), which
corresponds to the position of the star S\,1.

\subsection{Radio observations with GMRT}

We have mapped the low frequency radio continuum emission and 21\,cm \HI\
emission towards $\rho$ Ophiuchus using the upgraded Giant Metrewave Radio
Telescope \citep[uGMRT;][]{gupta2017}, India. The GMRT interferometer
comprises of 30 antennae, each of diameter 45\,m that are arranged in a
Y-shaped configuration \citep{Swarup1991}. Of these, twelve antennae are
located randomly within a central region of area $1\times1$\,km$^2$ and the
remaining eighteen antennae are placed along three arms, each of length
14\,km. The shortest and longest baselines are 105\,m and 25\,km respectively.
The configuration enables us to map large and small scale structures
simultaneously.  The observations were carried out during July 2018. The
radio source 3C286 was used as the primary flux calibrator and bandpass
calibrator whereas 1626-298 was used as phase calibrator. 

The observed field was centered at $\rho$ Ophiuchus ($\alpha_{J2000}$:
$16^h26^m34.0^s$, $\delta_{J2000}$: $-24^\circ23\arcmin28.0\arcsec$).
The radio continuum observations were carried out at 610  and 1420\,MHz.
The angular sizes of the largest structure observable with GMRT are
17$\arcmin$ and 7$\arcmin$ at 610 and 1420\,MHz, respectively. \HI\
observations were carried out along with the 1420\,MHz radio continuum
observations. The rest frequency of \HI\ line is 1420.4057\,MHz. The
\HI\ observations were performed with a bandwidth of 12.5\,MHz, which
was further divided into 8192 channels. The observing frequency was
estimated considering the LSR velocity of 3\,\kms\ \citep{pankonin1978}
as well as motions of the Earth and the Sun. The settings correspond to
a spectral resolution of 1.526\,kHz (velocity resolution of
0.322\,km/s). 

The data reduction was carried out using the NRAO Astronomical Image
Processing System (AIPS). The data sets were carefully checked and corrupted
data due to radio frequency interference, non-working antennas, bad
baselines etc. were flagged. After thorough flagging, the data was flux
and phase calibrated using the calibrators 3C286 and 1626-298. The data sets
were cleaned and deconvolved to create continuum maps.  Several iterations
of self-calibration were applied to minimize the phase errors. The final
images were then primary beam corrected. 


There are two sets of \HI\ observations: one on July 12 and the other
one on July 13, 2018. For the \HI\ observations, each of the final
calibrated data set was cleaned and deconvolved to produce a continuum
map.  Next, we subtracted the continuum (created from the line free
channels). The two data sets were then combined together to increase the
signal-to-noise ratio. We imaged the source with a UV tapering of
10~k$\lambda$ and a spectral cube was generated. The primary beam
correction was applied and the final image was obtained. The details of
the images are given in Table~\ref{radio_tb}.

\begin{table}
\begin{center}
\scriptsize 
\caption{Details of the radio observations with GMRT.}
\label{radio_tb}
\begin{tabular}{l c c c}
\hline\hline
 Frequency (MHz) & 610 &  \multicolumn{2}{c}{1420}\\
 & &Continuum&Line \\ 
 \hline
 Observation date &10 July 2018  &12, 13 July 2018  &12, 13 July 2018\\
 On source time (hrs) &3.9  &14.7 & 14.7\\
 Bandwidth (MHz) & 32 & 32&12.5\\
 Primary Beam & $45'.8$ & $19'.7$&\\
 Synthesized beam & 7\ptsec0 $\times$ 4\ptsec8  & 5\ptsec8 $\times$ 3\ptsec2 & 21\ptsec3
 $\times$ 14\ptsec3$^a$\\
 Position angle ($ ^\circ $) &$-5.4$  &30.9 &14.9\\
 Noise (mJy beam$^{-1}$) &1.5  &0.6 & 4.5\\
 \hline 
 \end{tabular}
\end{center}
 $^a$ Medium resolution opted to ensure detection of the emission with high fidelity
 in the channel maps
 \end{table}

\subsection{APEX}

The S\,1 PDR was mapped in CO(6--5) and in \thCO(6--5) using the SEPIA-660
receiver on the 12 m Atacama Pathfinder EXperiment (APEX)\footnote{APEX,
the Atacama Pathfinder Experiment is a collaboration between the
Max-Planck-Institut f\"ur Radioastronomie, Onsala Space Observatory
(OSO), and the European Southern Observatory (ESO).} telescope, located
at Llano de Chajnantor in the Atacama high desert of Chile
\citep{Gusten06}. The observations were part of the programme
m-0102.f-9524c-2018. SEPIA-660 is a SIS dual-polarization 2SB receiver
with an IF bandwidth of 4-12 GHz \citep{Belitsky2018}.  The backends
used are advanced Fast Fourier Transform Spectrometers \citep{Klein12}
with a bandwidth of 2$\times$4\,GHz and a native spectral resolution of
61\,kHz.  The rest frequencies  for CO(6--5) and \thCO(6--5) are
691.473076\,GHz and 661.0672766\,GHz, respectively. The HPBWs at
CO(6--5) and \thCO(6--5) are 9\ptsec0 and 9\ptsec4, respectively. The
main beam efficiency $\eta_{mb}$, is = 0.53, as measured from
observations of Jupiter (diameter 44.5\arcsec{})

The $^{12}$CO(6--5) map was observed on April 19, 2019  in on-the-fly
total power mode with an off position at 0\arcsec,+300\arcsec\
relative to S\,1 (RA $16^h26^m34.17^s$, Dec
$-24^\circ23\arcmin28.3\arcsec$). The weather conditions were good
(PWV 0.66\,mm) with a zenith optical depth of $\sim$ 0.75 resulting in
SSB system temperatures of $\sim$ 1000 K.  The map size was
235\arcsec $\times$200\arcsec, centered at (-17\arcsec5,0).  The field was
scanned in both RA and Dec with a spacing of 4\farcs5 (half the beam
size) and oversampled to 3\arcsec\ in scanning direction, resulting in
a uniformly sampled map with high fidelity.  Unfortunately, the off
position was not clean and we did a single point long integration
toward a far off at 0\arcsec,+1080\arcsec, which was used to correct
the map in the post processing stage.

On April 27, 2019 we observed a smaller map in \thCO(6--5) of the SW
part of the PDR, also in OTF TP mode. The map was centered at
(-75\arcsec,0) and  the map size was 60 $\times$ 120\arcsec, with the
same sampling strategy as above.  The weather conditions were good
(PWV 0.53\,mm) with a zenith optical depth of 0.67. The SSB system
temperature was $\sim$ 800 K.

The spectra were reduced in CLASS and calibrated in T$_{mb}$. 	We
removed a first order baseline and resampled the spectra to 0.5 km/s
velocity resolution.  The final data cubes (pixel size 9\arcsec) after
gridding have an rms main beam temperature noise per pixel of $\sim$
0.33\,K and 0.24\,K for CO(6--5) and \thCO(6--5), respectively.

\subsection{Auxiliary data}

For comparison with our observations, we have used maps of the $J$=3--2
transition of CO, $^{13}$CO and C$^{18}$O \citep{White15} and $J$=4--3
transition of HCO$^+$, all observed with JCMT using the HARP receiver
with a beamsize of 14\arcsec. The CO (and isotopologues) spectra were
observed as part of the Gould Belt Survey and the HCO$^+$(4--3) data
set, corresponding to the proposal M11AU13, was downloaded directly from
the JCMT archive at the Canadian Astronomical Data Centre (CADC). Both
these datasets have also been presented and compared with the previous
\CII\ observations of the S\,1 PDR  by \citet{mookerjea2018}. The
emission maps of S(2) and S(3) pure rotational transitions of H$_2$
observed with ISOCAM-CVF by \citet{larsson2017} were also used for
comparison.

\section{Results}

\subsection{Properties of the ionized gas around S\,1}

\begin{figure*}[h]
\centering
\hspace*{-1.5cm}\includegraphics[width=0.38\textwidth]{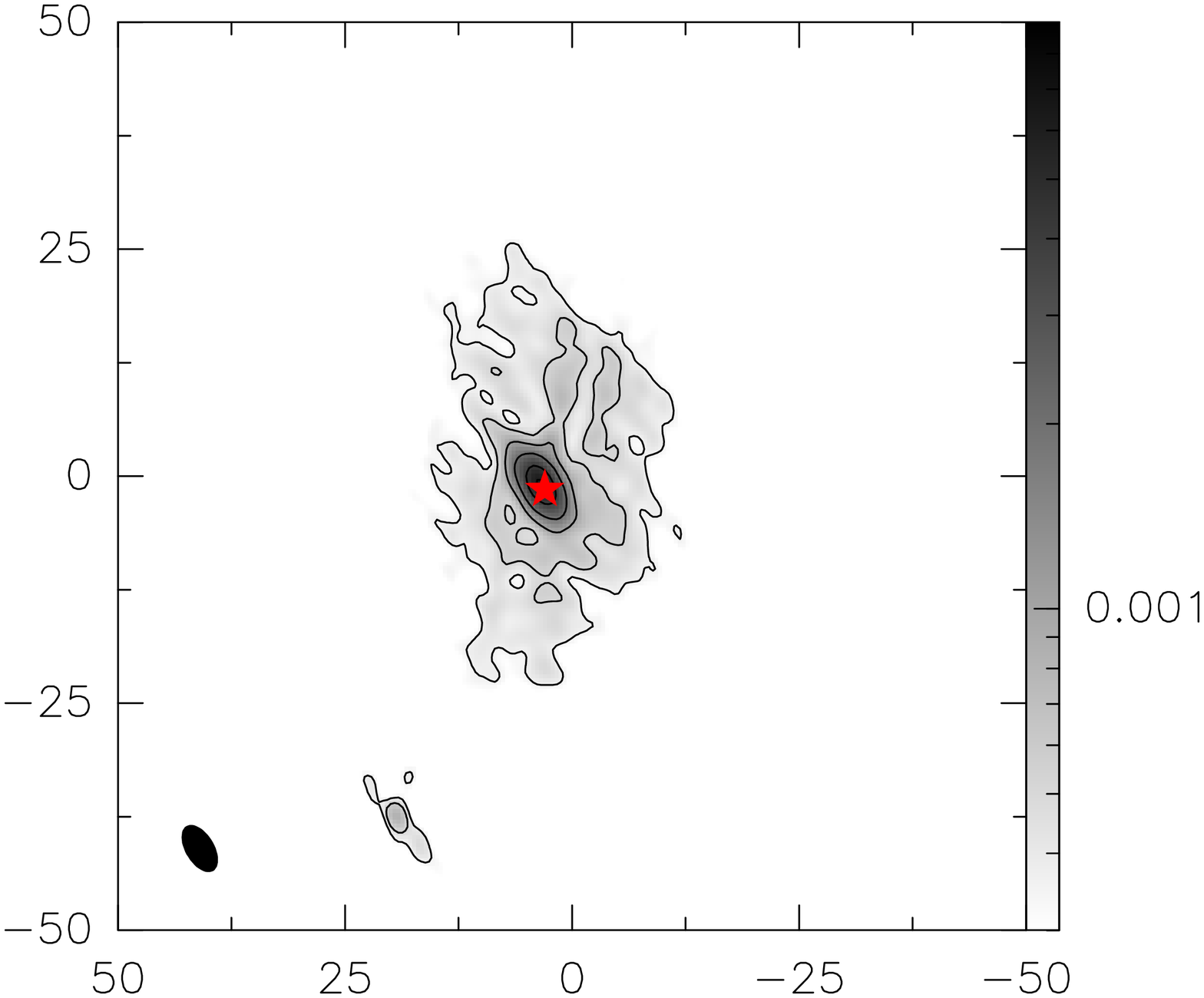} 
\hspace*{-1.2cm}\quad  
\includegraphics[width=0.38\textwidth]{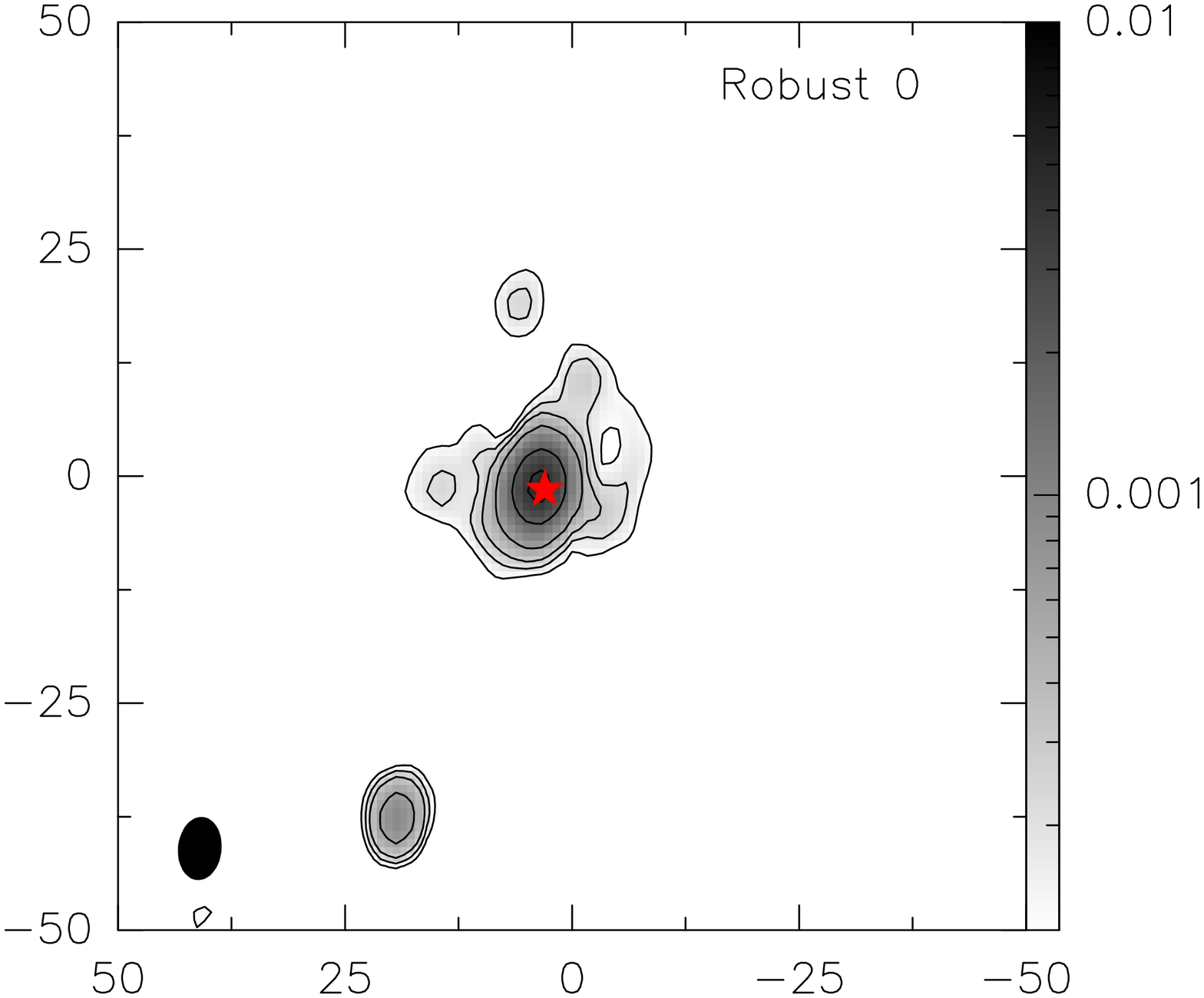} 
\hspace*{-1.2cm}\quad  
\caption{Radio continuum images of the region around S\,1 in
$\rho$\,Oph. (Left) At 1420\,MHz with contour at 0.36, 0.52, 1.0, 2.0,
4.0, 6.0\,mJy/beam. (Right) At 610\,MHz with contour at 0.12, 0.3, 0.6,
2.0, 4.0 and 6.0\,mJy/beam. The beamsizes are shown in the bottom left
corner of each panel. Asterisk marks the location of the embedded star
S\,1. The positional offsets (in arcseconds) are relative to the
center $\alpha$=16$^h$26$^m$34\fs175, $\delta$=$-24^\circ$ 23\arcmin\
28\farcs3 (J2000).
\label{fig_radiocont}}
\end{figure*}

The radio continuum emission from ionized plasma at 610 and 1420\,MHz
are shown in Fig.~\ref{fig_radiocont}. The emission at 610\,MHz shows a
bright core surrounded by low brightness diffuse emission. The emission
is extended up to 25\arcsec\ which corresponds to 0.02\,pc at a distance
of 137\,pc. The emission at 1420\,MHz extends up to 50\arcsec. A central
bright peak is observed surrounded by low surface brightness halo
emission. Both 610 and 1420~MHz images reveal an elongation in the North
west-South east direction, which is more prominent in the 1420\,MHz
image. Such elongated structures in radio emission are often indicative of
ionized jets from massive YSOs \citep[e.g.,][]{Purser2016}. The total flux densities at these frequencies are obtained using
a two component Gaussian fit to the emission. The flux density of the
central unresolved source is $5.6\pm0.2$\,mJy and that of diffuse halo
is $41.6\pm3.0$\,mJy at 1420\,MHz. The flux density of the central
source at 610~MHz is $6.8\pm2.0$\,mJy and that of diffuse emission is
$47.6\pm2.1$\,mJy.

Assuming that the diffuse emission at 1420\,MHz is optically thin, we
have
estimated the Lyman continuum photon rate and the spectral type of the star
responsible for ionized emission. The Lyman continuum photon flux at
1420~MHz towards S\,1 is estimated using the equation \citep{Schmiedeke2016}

\begin{equation}
\rm{\left[\frac{N_{Ly}}{s^{-1}}\right]=4.771\times10^{42}\left[\frac{S_\nu}{Jy}\right]\left[\frac{\nu}{GHz}\right]^{0.1}\left[\frac{T_e}{K}\right]^{-0.45}\left[\frac{d}{pc}\right]^2}
\end{equation}

where $\rm{S_\nu}$ is the flux density at frequency $\nu$ which is
41.59\,mJy at 1420\,MHz, $\rm{T_e}$ is the electron temperature which is
found to be 8200\,K based on electron temperature gradient across the
Galactocentric distance \citep{Quireza2006} and $d$ is the distance to the
source which is 137\,pc \citep{mookerjea2018}. Using the above expression,
the Lyman continuum photon rate is found to be $6.7\times10^{43}$~s$^{-1}$.
The estimated uncertainty in the electron temperature derived using the
formulation by \citep{Quireza2006} is $\approx 100$\,K. Thus, no additional
uncertainty in the estimated $N_{\rm Lyc}$ is introduced due to this. If a
single main sequence star is responsible for the ionization, then the
spectral type of the ZAMS star is earlier than B3V \citep{thompson1984}.
{\rm For a B3V star with $T_{\rm eff}$ = 18700\,K and R = 4.15\,\rsun,
Kurucz model \citep{Castelli2003} gives $N_{\rm Lyc}$ = 5.3$\times
10^{43}$\,s$^{-1}$  and for a B2V star with $T_{\rm eff}$ = 22000\,K and R
= 5.19\,\rsun, we get $N_{\rm Lyc}$ = 8.2$\times 10^{44}$\,s$^{-1}$. Thus
taking the uncertainties of the derived $N_{\rm Lyc}$ into account, we
conclude that the star S\,1 is most likely B2.5V or B3V, which is not
inconsistent with the SED fitting by \citet{mookerjea2018}. Although they
concluded a B4V type for the star, they noted that a B3V  would fit equally
well if a slightly larger extinction of 13.3 mag was adopted instead of
12.7\,mag.

From the VLA high frequency mapping of $\rho$ Oph at 5 and 15\,GHz,
\citet{andre1988} showed that the radio emission towards this region comes
from a non-thermal unresolved source surrounded by a thermal extended halo.
It is now known that S\,1 is a close binary \citep{Ortiz17} with the
secondary being responsible for the non-thermal emission.The flux density
of the central source at 1420\,MHz within the uncertainties is consistent
with the flux measurements of \citet{andre1988} and \citet{Stine1988}.
Using the flux densities of 6.8 and 5.6\,mJy at 610 and 1420\,MHz
respectively, we derive a spectral index of $-0.2\pm0.3$, where the
uncertainty in the derived index is contributed primarily by the
uncertainty of the 610\,MHz flux.

\subsection{The Morphology of the S\,1 PDR cavity}
\begin{figure}[h]
\begin{center}
\includegraphics[width=0.4\textwidth]{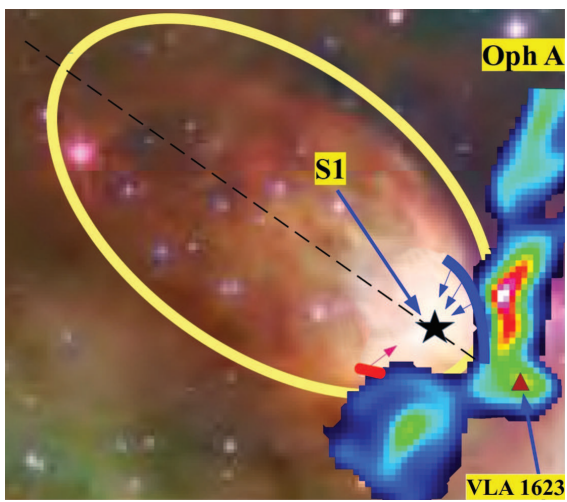}
\caption{Composite color image of the S\,1 PDR derived using the 3.6, 4.5
and 8\,\micron\ Spitzer observations \citep{Padgett08}. Also shown is
a cartoon of the morphology of the PDR that is derived in this paper on 
the basis of spatial and velocity dependence of emissions in multiple
tracers. The red and blue surfaces show respectively the red- and
blue-shifted fronts of the PDR. The color map corresponds to
N$_2$H$^+$ emission \citep{larsson2017} from the molecular cloud to the 
west. The yellow line demarcates the boundary of the PDR as identified
from our \CII\ observations and the 8\,\micron\ emission.
\label{fig_padgett}}
\end{center}
\end{figure}

IRAC and MIPS images \citep{Padgett08,Gutermuth09} show that
S\,1
illuminates a large elongated spheroidal or egg shaped cavity with the
major axis at a position angle of 54\degr\, and a length of $\sim$
10\farcm5  and a minor axis of $\sim$ 5\arcmin. There is very strong PDR
emission toward south-west (SW)  where it is blocked from expanding by
the surrounding dense molecular cloud. Toward the north-east (NE), where
the PDR shell emerges out of the cloud, the emission is rather faint and
barely visible. S\,1 is $\sim$80\arcsec\ from the SW tip of the PDR
shell\footnote{In the following we refer to NE to SW as being the
direction of the major axis of the PDR shell and north-west (NW) to
south-east(SE) being perpendicular to it.}. We have used a combination
of spatial and velocity information in the form of line integrated
emission, velocity-channel maps as well as position-velocity diagrams
along selected directions in the maps of the observed PDR tracers to
understand the basic geometry of the PDR associated with S\,1
(Fig.\,\ref{fig_padgett}).  To the west the PDR borders the Rho Oph A
ridge, which curves to the east south of the PDR. To the NW the
molecular cloud becomes very diffuse and is not seen in CO
\citep{White15}.  In the past there have been no spectral line
observations of the PDR emission east of S\,1 and most observations did
not even fully capture the PDR emission to N of S\,1, which prompted
\citet{larsson2017} to model the PDR shell as a gaseous sphere with a
radius of 80\arcsec. 

\begin{figure*}[h]
\centering
\includegraphics[width=0.78\textwidth]{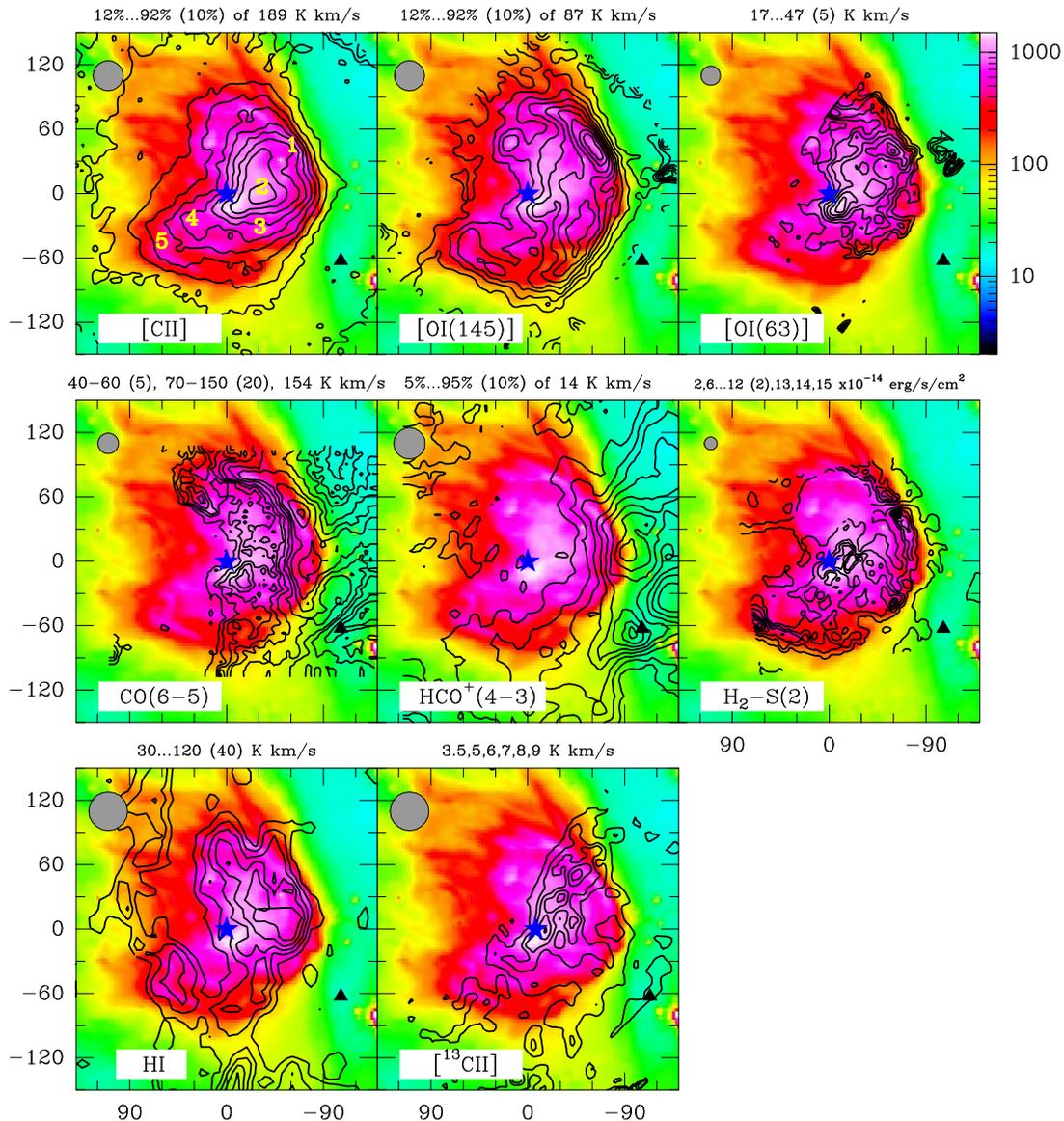}
\hfill
\caption{Comparison of 8\,\micron\ continuum image observed with
IRAC/Spitzer (color) with contours of integrated
intensity images of tracers (marked) overlayed on the 8\,\micron\
continuum image observed with IRAC/Spitzer.  The color scale is shown in
the wedge to the right of the top row, with numbers in units of MJy/sr.
For each of the tracers shown as contours, levels are shown on the top
of the panel and beams are shown at the top left corner of each panel.
The range of velocities over which integrations are done are as follows:
\CII\ -0.5 to 7.5\,\kms, \OI\ 145\,\micron\ 1 to 5.5\,\kms, \OI\
63\,\micron\ 0.5 to 6.5\,\kms, \HI\ -0.5 to 5\,\kms, CO(6--5) 0 to
10\,\kms, HCO$^+$(4--3) 0 to 10\,\kms\ and \thCII\ F=2--1 1 to 5\,\kms.  The 
positional offsets are
relative to the center $\alpha$=$16^h$26$^m$34\fs175,
$\delta$=$-24^\circ$ 23\arcmin\ 28\farcs3 (J2000). The asterisk and the
triangle, mark the positions of S\,1 and VLA\,1623, respectively. The
numbers
in the top left panel mark selected positions which are studied in
detail. The offsets for the positions are 1(-61,45), 2(-33,6), 
3(-31,-31), 4(32,-23) and 5(60,-45).
\label{fig_overlay}}
\end{figure*}

We have mapped extended regions around S\,1 in several PDR tracers: \CII,
\OI\ 63 and 145\,\micron, CO(6-5) and HCO$^+$(4-3). Although
HCO$^+$(4-3) is not really a PDR tracer and is mainly a dense gas
tracer, it does show some excess emission from the PDR
(Fig.\,\ref{fig_overlay}). These data have been compared with previous
observations of the $J$=3--2 transitions of CO, \thCO\ and \CeiO\
\citep{mookerjea2018}.  The emission seen in these PDR tracers match
well with the emission from neutral hydrogen (\HI) and other PDR tracers
such as 8\,\micron\ PAH emission and the S(2) pure rotational transition
of H$_2$ at 12.3\,\micron\ \citep[Fig.\,\ref{fig_overlay}; see
also][]{larsson2017}. The visual extinction toward S\,1 is $\sim$
13.3 mag and may even be  higher over part of the nebula
\citep{mookerjea2018}.
Additionally, the emission from all observed PDR tracers with the
exception of \OI\ 145 $\mu$m, which is optically thin, are heavily self
absorbed.  The \thCII\ F=2--1 line is also optically thin, but
relatively faint and only securely detected where the \CII\ emission is
strong.  The \OI\ 63 $\mu$m line is so strongly self absorbed, that no
emission is seen in the \vlsr\ range 3--4.5 \kms\
(Fig.~\ref{fig_chanmap1}). It is therefore unusable as a tracer of the
morphology of PDR cavity.  There is a strong CO(6--5) emission from
the Rho Oph A ridge. However, near the SW tip of the PDR cavity the
emission from the PDR starts to dominate. The CO(6--5) and \OI\
145\,\micron\ maps also capture the dense molecular PDR to the
north-east of S\,1.

\begin{figure*}[h]
\begin{center}
\includegraphics[width=0.95\textwidth]{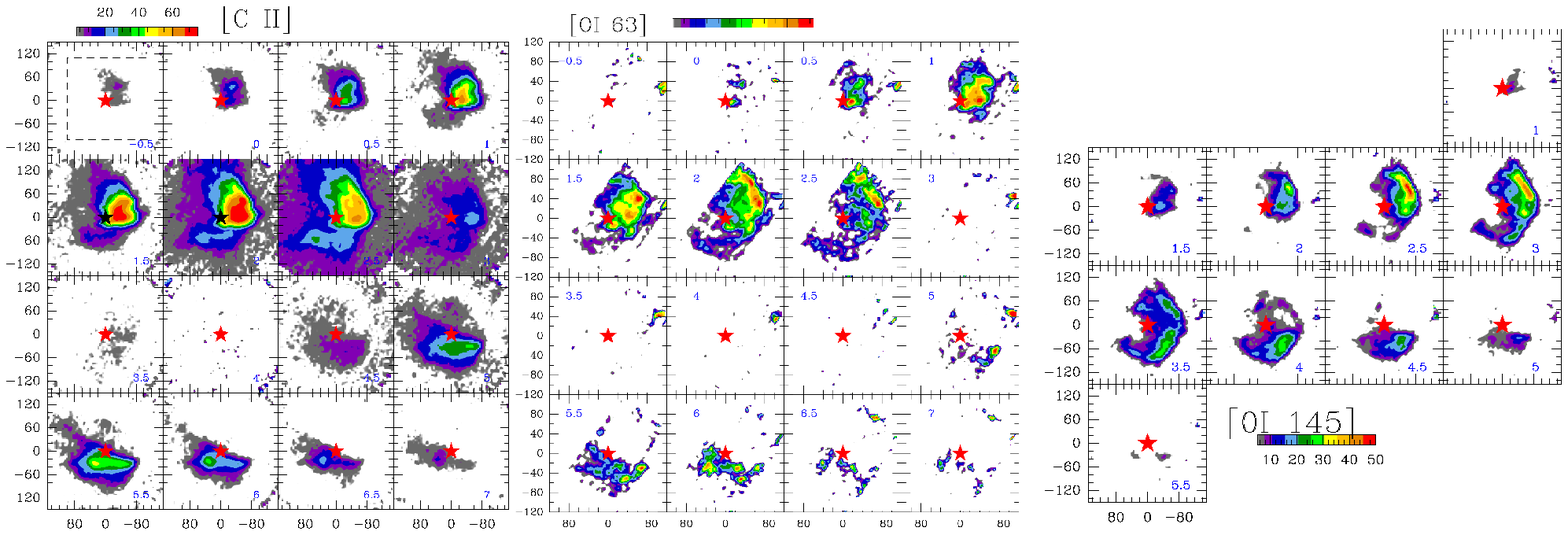}
\caption{Channel maps of \CII\ (left), \OI\ 63\,\micron\ (middle) and \OI\ 
145\micron\ (right).  The color scale for each map is shown next to the map. 
Velocities corresponding to the channel are marked in each panel. The red 
star sign marks the position of S\,1.  The positional offsets (in
arcseconds) are relative to the center $\alpha$=16$^h$26$^m$34\fs175, 
$\delta$=$-24^\circ$ 23\arcmin\ 28\farcs3 (J2000). The area mapped in
CO(6--5) is shown with dashed boundaries on the top-left panel of the \CII\
channel map.
\label{fig_chanmap1}}
\end{center}
\end{figure*}

\begin{figure*}[h]
\begin{center}
\includegraphics[width=0.95\textwidth]{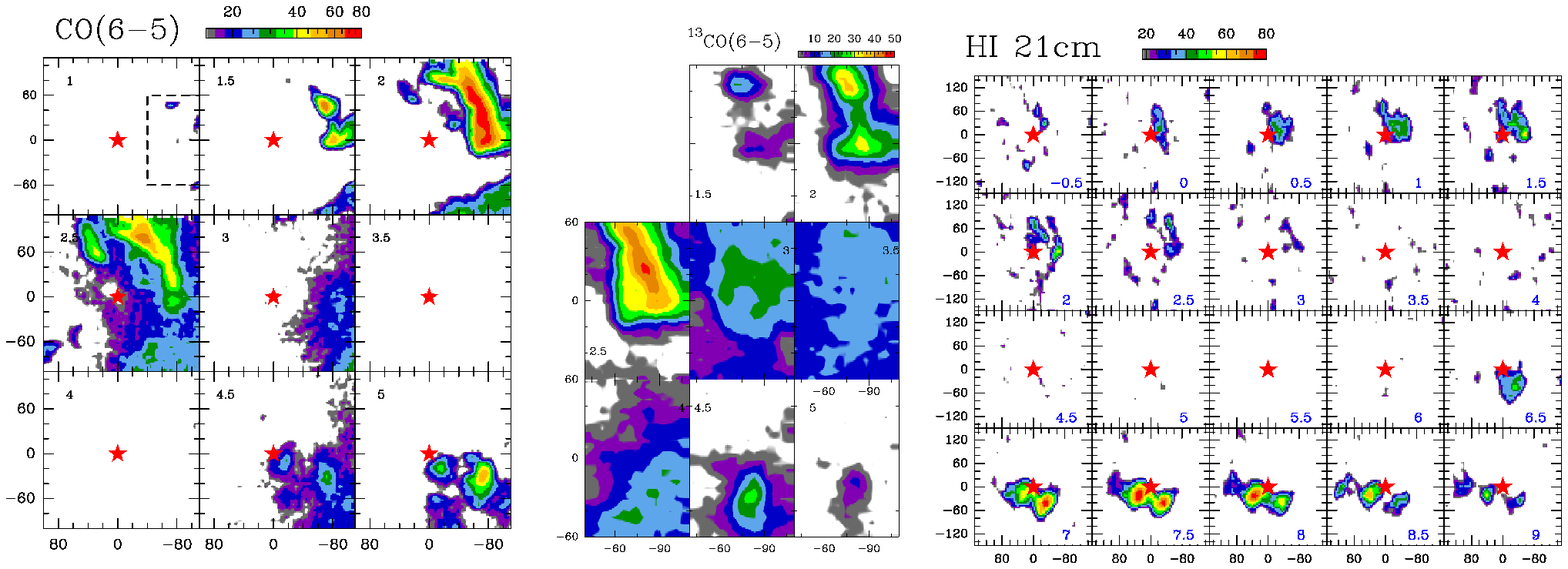}
\caption{Same as in Fig.\,\ref{fig_chanmap1} but for channel maps of
CO(6--5) (left), $^{13}$CO(6--5) (middle) and \HI\ (right)
emission. The area mapped in CO(6--5) is smaller than the maps in
Fig.\,\ref{fig_chanmap1}. The area mapped in $^{13}$CO(6--5) is marked with
dashed boundaries in the top left panel of the CO(6--5) channel map.
\label{fig_chanmap2}}
\end{center}
\end{figure*}

The channel maps of \CII\ and \OI\ 145\,\micron\
(Fig.\,\ref{fig_chanmap1}) show that the PDR emission is strongly
red-shifted on the south-eastern side of the nebula while the emission
is primarily blue-shifted on the north-western side. The \HI\ emission
is also strongly red-shifted on the south-eastern side. This implies
that gas is moving away from the observer on the south-eastern side,
while it is streaming toward the observer on the north-western side,
although  blue-shifted emission is also detected to the SE. At the SW
tip of the PDR one can see both blue- and red-shifted gas.  The
blue-shifted emission dominates to the NW, and red-shifted emission to
the SE. The same is true for the rest of the PDR nebula. This streaming
gas must be due to photo evaporating gas, which is commonly seen in
PDRs.  On the SE side of the cavity and toward SW, the PDR emission is
generally more red-shifted than the emission from the surrounding cloud.
Figure\,\ref{fig_chanmap2} shows that the CO(6--5) emission traces the
NW PDR boundary extremely well at velocities from 1.5--3 \kms\ and the
red-shifted filament south of S\,1. The CO(6--5) emission is strongly
self-absorbed in the PDR in the velocity range 3--4\,\kms, similar to
\OI\ 63\,\micron, though not as extreme. The emission from the filament
detected in CO(6--5) is by no means smooth.  It shows two clumps of
emission in the velocity channels from 4.5 and 5.5 \kms. The CO(6--5)
emission is quite faint  to the East of S\,1, where the \OI145\ and
\CII\ emission is still quite strong.  One can also see faint  CO(6--5)
emission NE of S\,1, which may be unrelated to the PDR.  The velocities
of emission of the \HI\ 21\,cm line are in general agreement with the
velocities and features traced by the PDR gas and not so much with
low-$J$ CO emission. The \HI\ emission is strongly affected by self
absorption between 3--6\,\kms\ and shows the east-west extended filament
but only at larger velocities of 6.5--9\,\kms\ than the PDR gas possibly
due to stronger self-absorption at lower velocities.
Figure\,\ref{fig_redblue} shows three-color composites of \CII\ and
\OI145\ emission for different velocity ranges.  This clearly shows that
the emission at near-cloud velocities (coded green) is close to S\,1,
while the blue-shifted emission arises from the north-west and the
red-shifted emission is to the south-east. The three-color plot of
\cplus\ additionally prominently shows a strong red-shifted filament
just SE of S\,1. 

\begin{figure}[h]
\begin{center}
\hspace*{-1.0cm}
\includegraphics[width=0.55\textwidth]{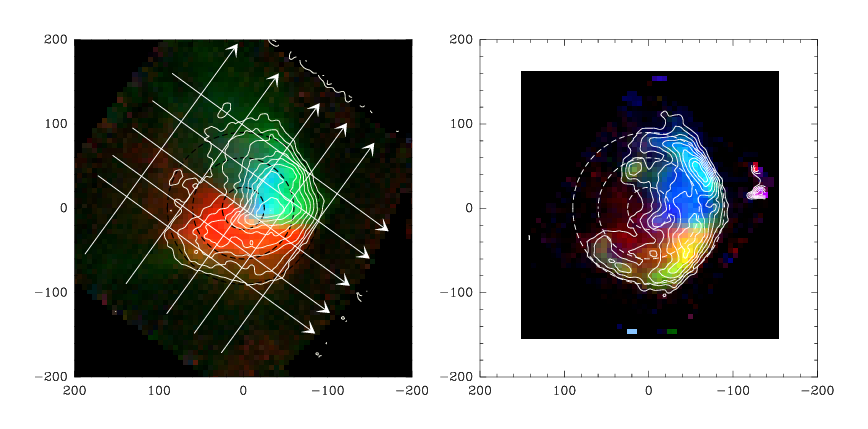}
\caption{Left: Three-color map of \CII\ emission in velocity interval \vlsr =
-0.5--1\,\kms, 1.5--3\,\kms and 4.5--7\,\kms\ shown as blue, green and
red channels respectively. The contours show the \CII\ emission in the
velocity interval -0.5--7\,\kms. The orthogonal sets of lines show 
the cuts along which position-velocity diagrams are derived. Right: Three-color
map of \OI\ 145\,\micron\ emission in the velocity intervals \vlsr =
1--2.5\,\kms, 3--4\,\kms\ and 4--5.5\,\kms\ as blue, green and red
channels respectively. The contours show the \OI\ 145\,\micron\ emission
in the velocity interval 1--5.5\,\kms.
\label{fig_redblue}}
\end{center}
\end{figure}

\begin{figure*}[h]
\begin{center}
\hspace*{-1cm}
\includegraphics[width=0.75\textwidth]{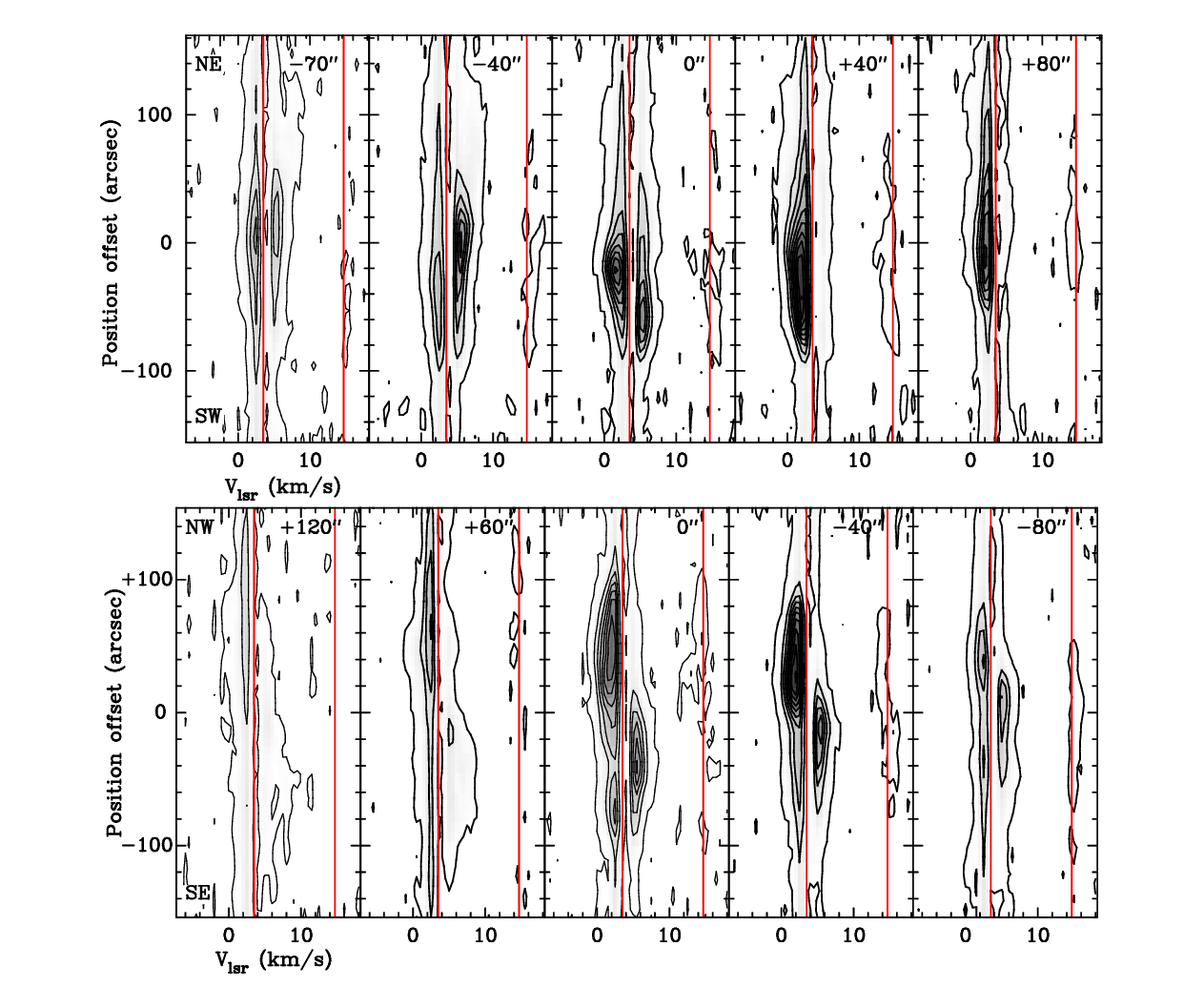}
\caption{Position-velocity diagrams of \CII\ emission along  the
parallel (top) and perpendicular (bottom) cuts  shown in
Fig.~\ref{fig_redblue}. The angular resolution is  18\arcsec. The two red
vertical lines mark the v$_{lsr}$ =+3.5\,\kms\ for \CII\ and  the
\thCII\  F=2--1 hyperfine line, which is offset by +11.2\,\kms\ in the
rest frame of \CII, or +14.7\,\kms.  The  contour levels are plotted
with linear step size and enhanced with gray scale from 1 K to 1.25
times peak temperature. For the parallel cuts 10 contours from 1--73\,K
with the peak temperature  at 73.1\,K are shown. The contour
levels for the perpendicular pv-plots are: 10 linear contours from
1--76\,K and gray scale from 1--96\,K. Offsets are relative  to
S\,1, which is at 0\arcsec. The parallel cuts go from NE to SW starting
from the South. The perpendicular cuts go from NW to SE and start East
of S\,1.
\label{fig_ciipv}}
\end{center}
\end{figure*}

\begin{figure*}[h]
\hspace*{-1cm}
\begin{center}
\includegraphics[width=0.75\textwidth]{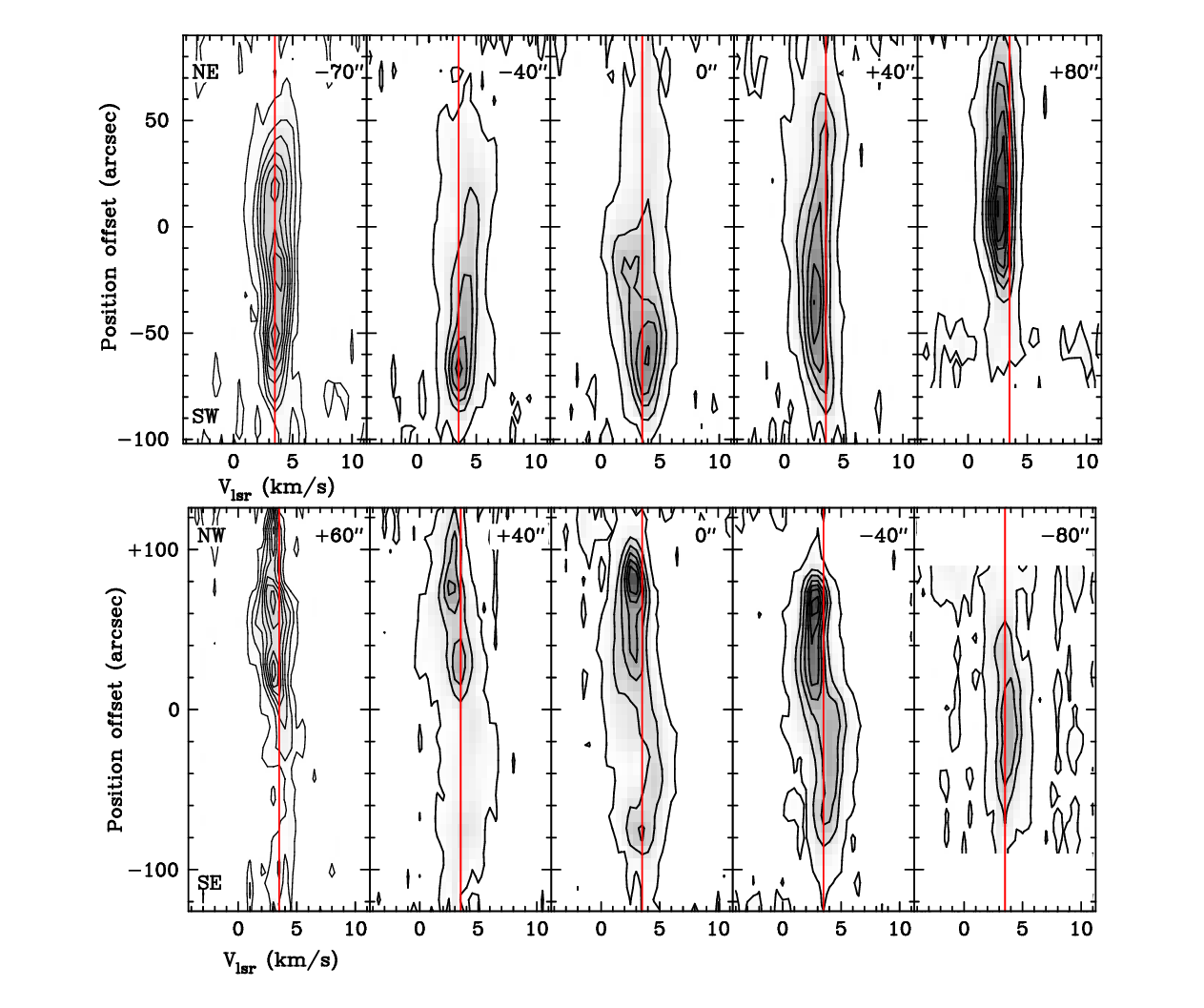}
\caption{Position-velocity diagrams of \OI\ 145 $\mu$m emission along
the same parallel (top) and perpendicular (bottom) cuts  as \CII\ in
Fig.~\ref{fig_ciipv}. The angular resolution is the same as for \CII,
i.e., 18\,\arcsec. Eight linear contours going from 1.25--42\,K are
shown, while the gray scale is between 1.25--52.9\,K.
\label{fig_oipv}}
\end{center}
\end{figure*}

\begin{figure*}[h]
\hspace*{-2cm}
\begin{center}
\includegraphics[width=0.75\textwidth]{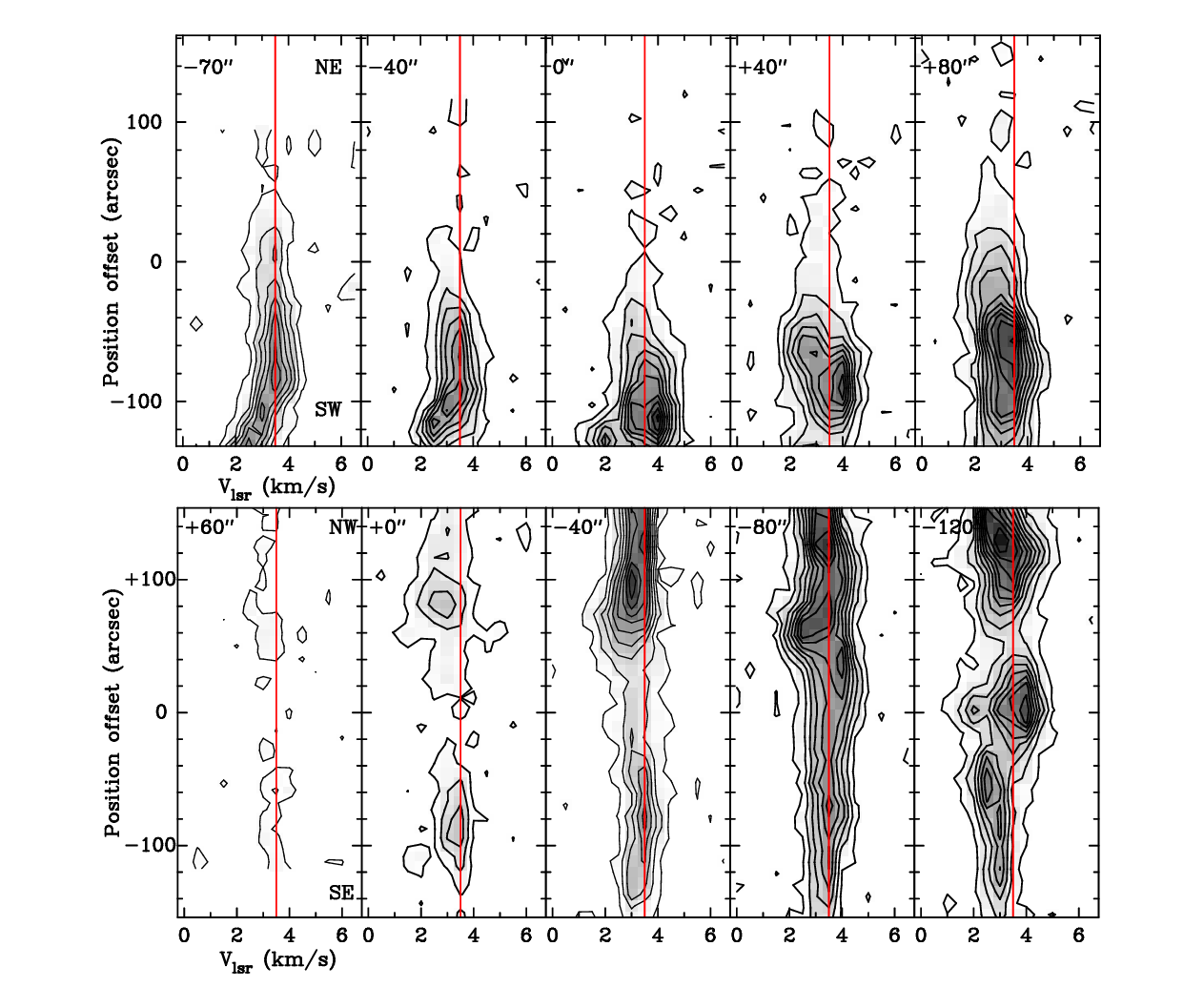}
\caption{Position-velocity diagrams of HCO$^+$(4--3) emission along
parallel (top) and perpendicular (bottom) cuts shown in
Fig.\,\ref{fig_redblue} and plotted the same way as Figs.~\ref{fig_ciipv} \&
\ref{fig_oipv} using the same angular resolution as for \CII. Ten linear
contours  from 0.3--6.9\,K are shown and thee peak intensity is 7.0\,K.
To the NE, the HCO$^+$ is barely detectable 60\arcsec\ from
S\,1 confirming
that the density of the surrounding cloud falls off towards the NE.  The
southwestern most perpendicular position velocity cut 120\arcsec\ west of
S\,1 does not cross the PDR, but is plotted to illustrate that the emission
in the Rho Oph A ridge is more blue-shifted that the emission
from the S\,1
PDR.
\label{fig_hcoppv}}
\end{center}
\end{figure*}

\begin{figure*}[h]
\hspace*{-2cm}
\begin{center}
\includegraphics[width=0.75\textwidth]{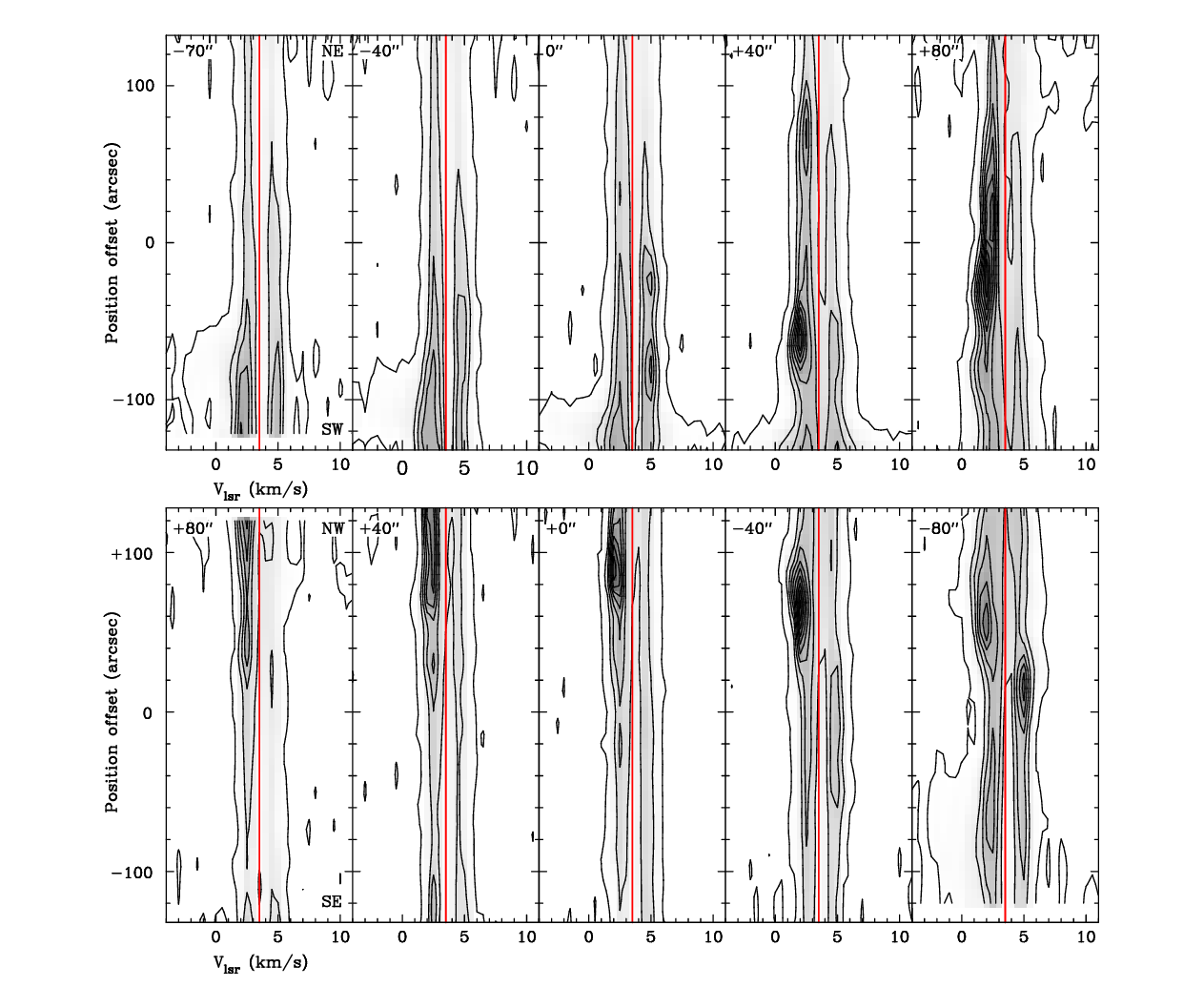}
\caption{Position-velocity diagrams of CO(6--5) emission along parallel
(top) and perpendicular (bottom) cuts shown in Fig.\,\ref{fig_redblue} and
plotted the same way as Figs.~\ref{fig_ciipv} \& \ref{fig_oipv} using the
same angular resolution as for \CII. Ten linear contours  from
0.8--68\,K are shown and the gray scale goes from 0.8--85\,K.  In the
parallel cuts red- and blue-shifted emission from the VLA\,1623 outflow in
the SW are observed. The blue-shifted outflow lobe in the SE is detected
in the perpendicular cut 80\arcsec\ from S\,1.  This bipolar outflow is
unrelated to the S\,1 PDR. There is no evidence for PDR emission in the
perpendicular cut 80\arcsec\ NE of S\,1. The CO(6--5) emission is faint and
most likely from the surrounding molecular cloud.
\label{fig_co65pv}}
\end{center}
\end{figure*}

In order get a more detailed view of the S\,1 PDR cavity, we created
position velocity diagrams (Figs\,\ref{fig_ciipv}--\ref{fig_co65pv}) of
several PDR tracers both along cuts parallel and perpendicular to the
major axis of the cavity as shown in Fig.\,\ref{fig_redblue}. In the
following, we refer to these position-velocity diagrams as pv-cuts. The
\CII\ map is more extended than the other maps  and in \CII\ one can
still see faint PDR emission 120\arcsec\ to the NE of S\,1
(Fig.\,\ref{fig_ciipv}). There will be \CII\  emission even further to
the East, although it is likely to be faint. However, since the gas
densities and FUV radiation field is low, it is possible that only the rim
outlining the PDR shell can be detected.  The position-velocity diagrams
show that the SW tip of the PDR is somewhat red-shifted suggesting that
the PDR is tilted away from us in the SW and coming toward us in the NE.
The perpendicular pv-cuts across the PDR cavity show that the cavity is
less extended to the SE ($\sim$ 80\arcsec{}) than to the NW ($\sim$
100\arcsec{}), suggesting that the surrounding molecular cloud must be
much denser on the SE side than on the NW side, which slows down the
expansion of PDR cavity to the SE compared to the NW side. The
perpendicular pv-cuts also show strong blue-shifted emission to the NW
with more red- than blue-shifted emission to the SE.  The flip from
red-shifted  to blue-shifted emission occurs roughly at the symmetry
axis of the PDR cavity.  The HCO$^+$ emission, which is dominated by the
surrounding cloud, also shows the same velocity gradient across the PDR
cavity. These pv-cuts suggest that on the SE side of the cavity the
surrounding cloud must be very dense on the front side, i.e. the side
facing us, while on the NW side the cloud is denser on the back side.
This forces the photo evaporation flow from the PDR to be mostly
red-shifted in the SE and blue-shifted in the NW. We have tried to
visualize this in the cross-sectional view of the PDR SW of
S\,1
(Fig.\,\ref{fig_padgett}), which is a  very simplified picture,
because the PDR layer is by no means smooth.  There may be ridges and
valleys and there can also be dense clumps of gas inside the PDR, like
what appeared to be the case for NGC2023 \citep{Sandell15}. There is a
strong red-shifted filament just SE of S\,1, which stands out
prominently in the \OI\ 63 $\mu$m channel maps at velocities from 5 --
6.5 \kms\ (Fig.~\ref{fig_chanmap1}). This filament is also seen in the
\OI\ 145\,$\mu$m, \CII, \HI\ and CO(6--5) channel maps
(Fig.~\ref{fig_chanmap1} and \ref{fig_chanmap2}).

\subsection{Analysis of Spectral Profiles}

The \CII, CO(6--5), \HI\ spectra throughout the observed map show
self-absorption, along with red- and blue-shifted line wings. The
\CII\ lines are the broadest among the PDR and molecular tracers. The
\OI\ 145\,\micron\ spectra typically show a single peak centered on
the absorption in the \CII\ spectra as well as an extended blue wing
at a few positions.  The \OI\ 63\,\micron\ on the other hand is
completely absorbed between 3 -- 4.5\,\kms.  The spectra of
\thCO(6--5) show absorption dips at positions close to the CO(6--5)
peak but is single-peaked at positions to the south-west of the map as
well as in the region immediately to the west of the S\,1. This suggests
that the foreground material is optically thin in \OI\ 145\,\micron\
throughout the entire map, while \thCO(6--5) is still moderately optically thick
at a few positions.  Figure\,\ref{fig_selspec} shows the spectra at
the location of S\,1 and five additionally selected positions (shown in
Fig.\,\ref{fig_overlay}) within the mapped region, which sample a
variety of \CII\ lineshapes that are representative of the entire map. 

\begin{figure*}[h]
\begin{center}
\includegraphics[width=0.95\textwidth]{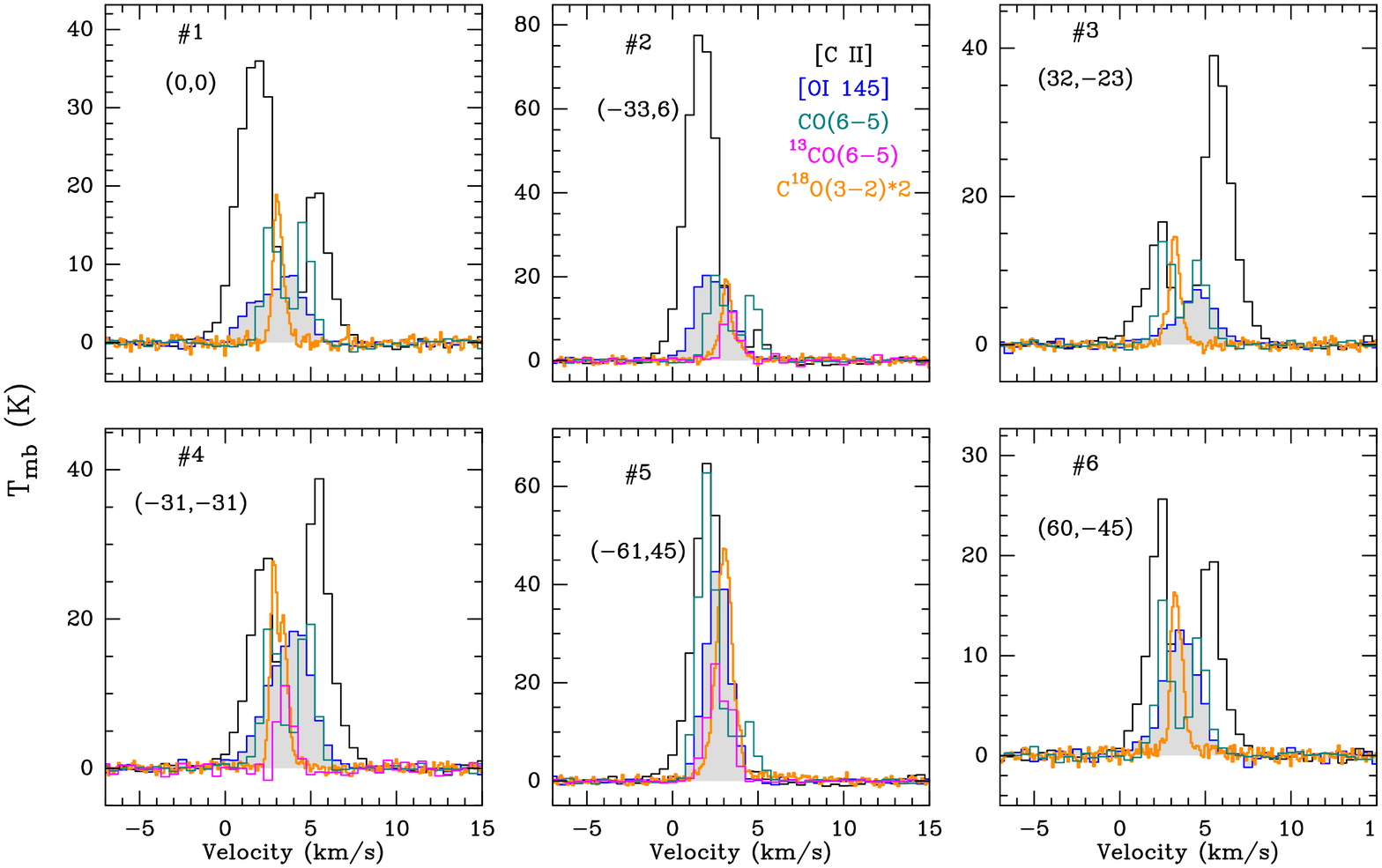}
\caption{Comparison of spectra of \CII, \OI, \CeiO(3--2), CO(6--5) and 
\thCO(6--5) transitions at the selected positions. The positions are shown in
Fig.\,\ref{fig_overlay}. The dataset corresponding to the \CeiO(3--2) spectra 
shown here were presented by \citet{mookerjea2018}.  All spectra shown
have been co-added across a 20\arcsec\ field.
\label{fig_selspec}}
\end{center}
\end{figure*}

Our observations are deep enough to detect the $F$ = 2--1 transitions of
\thCII\ at 90 positions in the entire region.  Closer inspection of the
\CII\ spectra at five out of the six selected positions also reveal
clear detection of the $F$ = 2--1 transition of \thCII. Unlike the \CII\
profiles which show a strong absorption dip, the \thCII\ spectra peak exactly at
the location of the \CII\ dip (Fig.\,\ref{fig_slab}).  Similar trends are also
visible when the \OI\ 63 and 145\,\micron\ spectra are compared at these five
positions (Fig.\,\ref{fig_oislab}).

\begin{figure}[h]
\begin{center}
\includegraphics[width=0.45\textwidth]{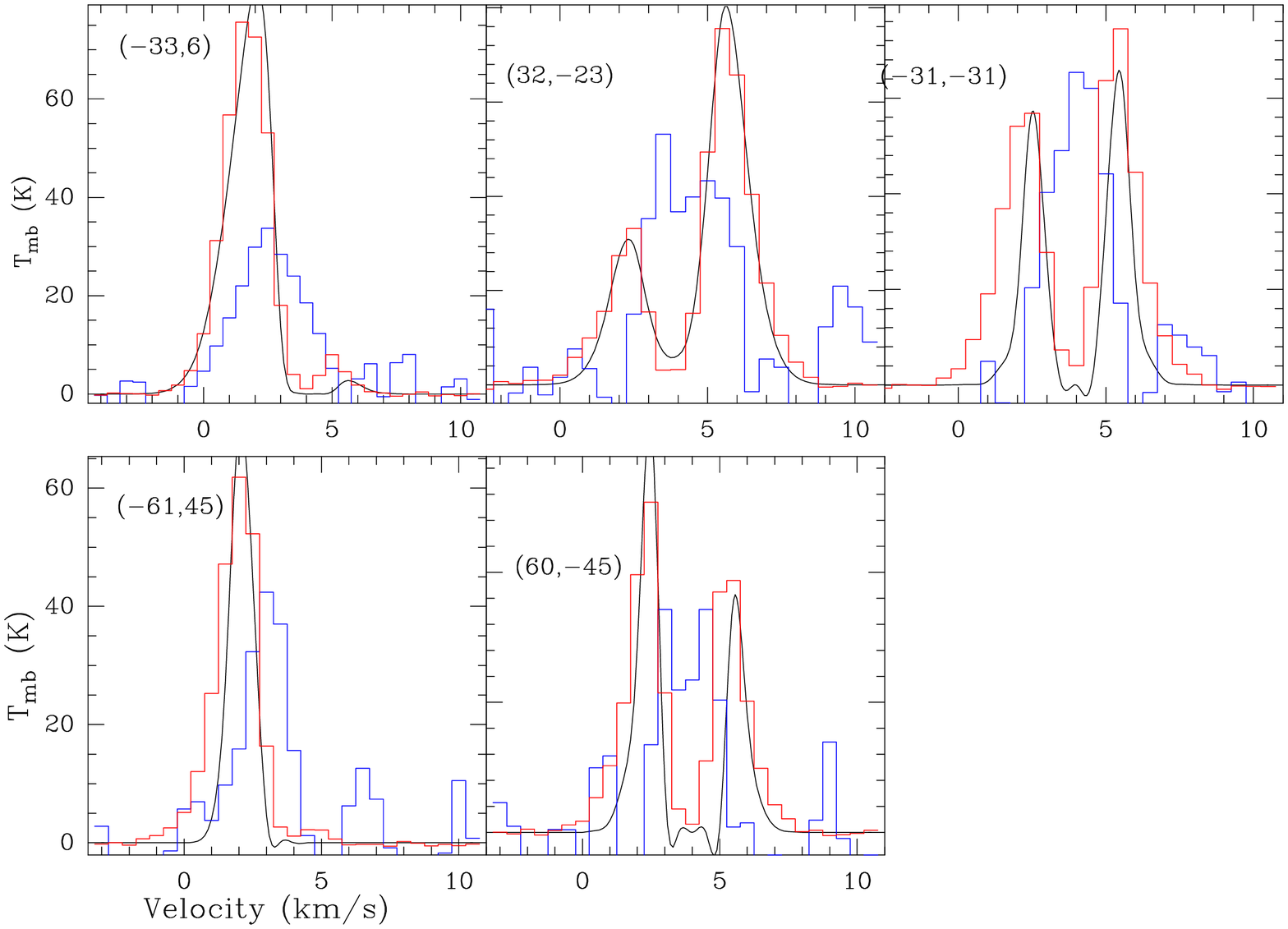}
\caption{Comparison of \CII\ (red) and \thCII\ (blue) spectra averaged over
20\arcsec\ with centers at offsets shown in the panels. The \thCII\
spectra have been scaled by a factor of 20 to make the spectra visible 
alongside \CII. The smooth curve (black) shows the fit to \CII\ spectrum 
obtained by attenuating the scaled \thCII\ spectrum by the absorption due 
to the foreground cloud.
\label{fig_slab}}
\end{center}
\end{figure}

\begin{figure}[h]
\begin{center}
\includegraphics[width=0.45\textwidth]{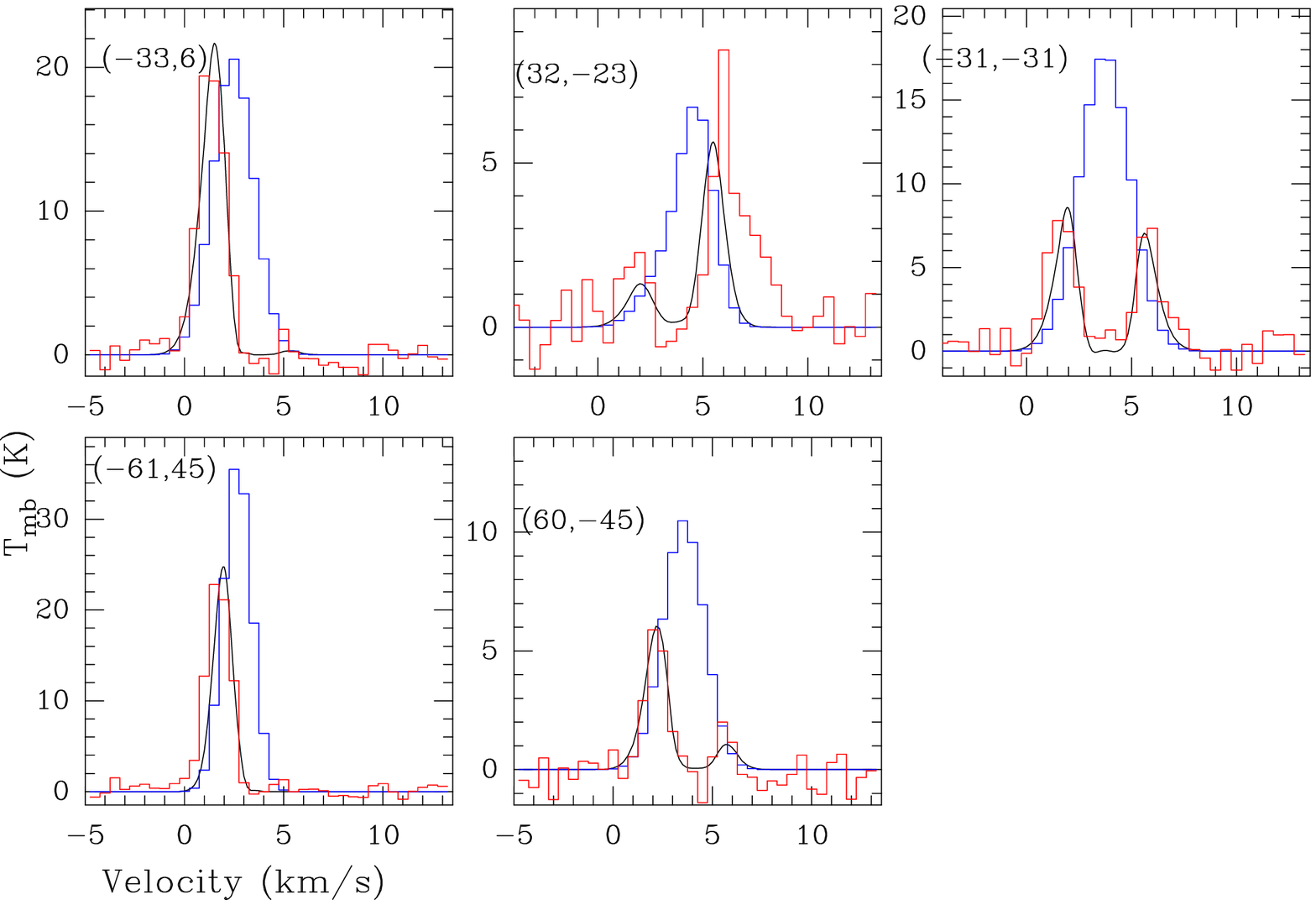}
\caption{Comparison of \OI 63 (red) and \OI\ 145\,\micron\ (blue) spectra 
co-added over 14\arcsec\ with centers at offsets shown in the panels. The 
smooth curve (black) shows the fit to the \OI 63 spectrum obtained by 
attenuating the scaled (by a factor 2 corresponding to the typical ratio between 
the two \OI\ lines in temperature units) \OI\ 145\,\micron\ spectrum by 
absorption due to foreground material.
\label{fig_oislab}}
\end{center}
\end{figure}

We start with trying to  understand the role of the foreground material in shaping the
\CII\ and \OI\ 63\,\micron\ spectral profiles characterized by deep self absorption.
We model the observed \CII\  and \OI\ 63\,\micron\ spectra considering two layers, a
background one (the main PDR) that emits and a foreground cloud that absorbs. We take
an empirical approach and assume that the \thCII\  and \OI\ 145\,\micron\ spectra
which are optically thin, represent (except for a {\em scaling factor}) the \CII\ and
the \OI\ 63\,\micron\ spectra respectively, that the PDR would emit in the absence of
the foreground absorbing gas. The scaling factor used for the \thCII\ spectrum is
$^{12}$C/$^{13}$C=70  (after correcting for the detection of only the strongest of the
hyperfine structure line of \thCII\ which accounts for 0.625 of total \thCII\
intensity).  The ratio of the \OI63/\OI145\ intensities in the background emitting gas depends
strongly on the physical conditions, here we use a scaling factor of 2 for the \OI\
145\,\micron\ spectrum which corresponds to  $F_{63\,\micron}$/$F_{145\,\micron}$ of
24, a value that is representative for warm, medium dense PDR conditions \citep[e.g.
Fig.  5]{Goldsmith2019}.  Thus, we consider that at each position the \CII\ (\OI\
63\,\micron) spectrum is given by the scaled \thCII\ (\OI\ 145\,\micron) spectrum and
the foreground gas is a pure absorbing screen with a constant velocity, width and
optical depth.  In this approach we assume that the effect of the foreground material
on the background emission is to attenuate the latter by the factor
$\exp\left({-\tau_0\exp[-4\ln 2\,(\upsilon-\upsilon_0)^2/\Delta \upsilon^2]}\right)$.
Here $\tau_0$, $\upsilon_0$ and $\Delta \upsilon$ denote the peak optical depth, the
velocity at which the foreground cloud has a value of $\tau_0$, and the FWHM of the
foreground absorption profile.  In order to generate the \thCII\ spectra with higher
fidelity we have co-added the \thCII\ and all other spectra over 20\arcsec\ with
centers at the selected offsets. Additionally, to improve the quality of the template
spectra generated from \thCII,  and \OI\ 145\,\micron, we do not use the observed
spectra to simulate the emission, but instead we use the Gaussian profiles generated
from fits to the observed spectra.  Figures\,\ref{fig_slab} and \ref{fig_oislab} also
show the results of such two-slab modeling at five selected positions in the observed
region.

\begin{table}[h]
\caption{The peak optical depth, the corresponding velocity and the FWHM
of the foreground absorbing gas derived from modeling the \CII\ and 
\OI 63\,\micron\ spectra with a two-slab model. The ${\rm
T_{kin}^{min}}$ presented is an estimate of the lower limit of 
kinetic temperature of the \CII\ emitting gas. It corresponds to the 
peak of the fitted background \CII\ spectrum, assumed to have an optical 
thickness of unity. \label{tab_twoslab}}
\scriptsize
\begin{tabular}{rrrrrrrr}
\hline
Position & Species & Ratio & $\tau_0$ & $\upsilon_{\rm LSR}$ &
$\Delta \upsilon$ & $N_{\rm abs}$ & \tkin\\
& & & & \kms & \kms & \cmsq & K\\
\hline
\hline
(-61,45)  & \CII   & 70 &  4.7 &  3.5 &  1.4 & 9.9\,10$^{17}$ & 157\\
          & \OI 63 &  2 &  6.8 &  4.0 &  2.2 & 3.0\,10$^{18}$ &   \\
\hline
(-33, 6)  & \CII   & 70 &  3.5 &  4.4 &  3.4 & 1.8\,10$^{18}$ & 121\\
          & \OI 63 &  2 &  7.9 &  3.7 &  1.9 & 3.0\,10$^{18}$ &   \\
\hline
(-31,-31) & \CII   & 70 &  4.8 &  3.9 &  1.5 & 1.1\,10$^{18}$ & 130\\
          & \OI 63 &  2 &  8.5 &  3.8 &  1.7 & 2.9\,10$^{18}$ &   \\
\hline
(32,-23)  & \CII   & 70 &  3.5 &  3.9 &  2.2 & 1.2\,10$^{18}$ & 97\\
          & \OI 63 &  2 &  4.1 &  3.8 &  1.9 & 1.6\,10$^{18}$ &   \\
\hline
( 60,-45) & \CII   & 70 &  4.6 &  3.9 &  1.4 & 9.7\,10$^{17}$ & 79\\
          & \OI 63 &  2 &  6.0 &  4.1 &  1.9 & 2.3\,10$^{18}$ &   \\
\hline
\hline
\end{tabular}
\end{table}

Table\,\ref{tab_twoslab} presents the results of fitting of the
two-component model to the \CII\ and \OI 63\,\micron\ spectra. The
\cplus\ absorbing layer shows peak opacities ($\tau_0$) between
3.5--4.8 at velocities ($\upsilon_0$) between $\sim
3.5$--4.4\,\kms, with a width ($\Delta \upsilon$) between
1.4--3.4\,\kms. The O$^0$ absorbing layer gives rise to peak
opacities for \OI 63\,\micron\ of 4--8, at velocities between
3.7--4.1\,\kms\ with a linewidth of 1.7--2.2\,\kms.

We point out that in particular the blue part of the \CII\ spectrum is not
fully reproduced at some of the positions, likely due to the fact that the
\thCII\ lines being fainter do not trace the line wings where \CII\ is
optically thin. Similarly, at (32,-23) for the fit to the \OI\
63\,\micron\ spectrum, a broader red-shifted velocity component is
completely missed out and part of the blue wing is not fully reproduced by
the fit based on the \OI\ 145\,\micron\ spectra. For both (-31,-31) and
(-61,45) the fits lack somewhat at lower velocities.  The  central
velocity and linewidth of the foreground absorbing component derived for
fits to both \CII\ and \OI\ 63\,\micron\ spectra are consistent with the
two-component LTE-based modeling that was performed by
\citet{mookerjea2018}. However, the fit presented here is better
constrained because of the availability of \thCII\ and \OI\ 145\,\micron. 

Table\,\ref{tab_twoslab} presents the column densities in the lower energy
level of \cplus\ and O$^0$ in the foreground absorbing gas estimated based
on the $\int \tau\, d\upsilon$ values derived from the fits and using the
relation:

\begin{equation}
N_l = \frac{g_l}{g_u}\frac{8\pi}{\lambda^3A_{ul}}\int{\tau_{ul}\, d\upsilon}
\end{equation}  

where, $g_{\rm l}$ and $g_{\rm u}$ denote the statistical weights of the
lower and upper energy levels, $A_{\rm ul}$ denotes the Einstein's
A-coefficient for spontaneous emission and $\lambda$ denotes the
wavelength of the transition. 

We obtain $N$(O) of the absorbing gas to be between
(2.3--3.0)\,10$^{18}$\,\cmsq.  For these column densities,  based on
non-LTE calculations using RADEX \citep{vdtak2007}, over a large
range of temperatures (20--300\,K) and volume densities
(10$^4$--10$^7$\,\cmcub) the \OI\ 145\,\micron\ line is optically thin.
This is also consistent with our observations. Since the $T_{\rm A}$
ratio for the two \OI\ lines depends on the physical conditions, there is
significant uncertainty in the derived values of the peak optical depth,
although the fitted central velocity and linewidth are fairly robust
against the assumed scale factor. Further, the estimate of $N$(O$^0$)
assumes that all O atoms in the absorbing layer are in the ground $^3$P$_2$
level, which is reasonable since \OI\ 145\,\micron\ is not seen in
absorption. We estimate the column density of \cplus\ in the absorbing gas
to be between (1--2)\,10$^{18}$\,\cmsq. We note, that the estimate of
$N$(\cplus) assumes that more than 95\% of the \cplus\ are in the ground
state. For excitation temperatures of foreground gas higher than 25\,K, the
derived $N$(\cplus) is a lower limit. Thus the $N$(O$^0$)/$N$(\cplus) ratio
in the foreground absorbing layer is between 2--3. 

This toy model confirms that both the \CII\ and \OI 63\,\micron\ spectra
are strongly self-absorbed by the foreground PDR gas.  Additionally, based
on the derived velocity at which the peak optical depth occurs for both
species ($>$3.2\,\kms), we also find that the absorption occurs due to
the temperature gradient in the PDR gas itself and not due to the ambient
molecular gas traced by the $J$=3--2 transitions of CO and its
isotopologues.  


\subsection{Temperature of \CII\ emitting PDR gas}

In our analysis of the absorption features in the \CII\ profile, we have
considered the background \CII\ profile to be an optically thin
scaled-up version of the \thCII\ profile. However, based on the recent
\CII\ observations of most of the Galactic PDRs it is likely that the
background PDR emission is likely to have an optical depth close to 1.
\citet{Guevara2020} performed an elaborate fitting procedure involving
multi-component LTE components to explain optically thick background
\CII\ spectra that are absorbed by a foreground layer of gas. Here we
approximate the background \CII\ spectrum, by assuming it to be a scaled
up version of the \thCII\ spectrum modulated by an optical depth of 1.
The Planck-corrected peak of the optically thick \CII\ spectrum so derived
provides a lower limit of the temperature of the \CII\ emitting PDR gas.
Table\,\ref{tab_twoslab} also presents the temperatures of the \CII\
emitting PDR gas at the selected positions estimated using this method.

We use the integrated line intensities of the optically thin \thCII\
spectra at these selected positions to estimate $N$(\cplus).  We estimate
the total integrated intensity of \thCII\ by considering that the observed
$F$=2--1 transition accounts for 62.5\% of the total intensity
(Table\,\ref{tab_gauss}). The column density of \thcplus\ is estimated
following Eq (26) from \citet{goldsmith2012}:

\begin{equation}
N(^{13}{\rm C^+}) = {\rm \frac{8\pi k_{\rm B}\nu_{ul}^2}{A_{ul}hc^3}\left[1+0.5{\rm
e^{91.25/T_{kin}}}\left(1+\frac{A_{ul}}{C_{ul}}\right)\right] \int T_{\rm mb}d\upsilon}
\end{equation}

where, $\nu_{ul}$ = 1900.4661\,GHz, $A_{ul}$ = 2.3$\times 10^{-6}$\,s$^{-1}$,
\tkin\ is the gas kinetic temperature, the collision rate is $C_{ul}$ = $R_{ul}n$
with $R_{ul}$ being the collision rate coefficient with H$_2$ or H$^0$ which
depends on $T_{\rm kin}$ and $n$ is the volume density of H. For $n_{\rm
H}>10^4$\,\cmcub, $C_{ul}\gg\,A_{ul}$ so that the last term in Eq. 3 can be
neglected.

We assume a $^{12}$C/$^{13}$C ratio of 70, based on  the Galactocentric
distance of the S\,1 PDR \citep{Wilson1994}, using the observed integrated
\thCII\ intensities and the estimated \tkin\ (Table\,\ref{tab_twoslab}), we
estimate $N$(\cplus) at the selected positions to be between
1.3--3.8$\times 10^{18}$\,\cmsq\ (Table\,\ref{tab_gauss}). Comparing the
total \cplus\ column density derived here with $N$(\cplus) estimated for
the foreground absorbing gas (Table\,\ref{tab_twoslab}), we find that for
all positions the column density of the colder foreground gas is
approximately one-third of the $N$(\cplus) of the background PDR gas.

\begin{table}[h]
\scriptsize
\caption{Results of fitting Gaussian profiles with single velocity
components to observed spectra at selected positions. $N$(\cplus) 
is calculated assuming the lower limits of kinetic temperature, $T_{\rm kin}^{\rm min}$ 
from Table\,\ref{tab_twoslab}.\label{tab_gauss}}
\begin{tabular}{rlrrrr}
\hline
\hline
($\Delta \alpha$,  $\Delta \delta$) & Transition &  $I$\phn\phn\phn\phn &
$\upsilon_{\rm LSR}$\phn\phn\phn & $\Delta \upsilon$\phn\phn\phn  &
$N$(\cplus)\\
& & K\,\kms & \kms\phn\phn & \kms\phn & \cmsq\\
\hline
(-61,45)  & \thCII      &   3.99$\pm$0.38 & 2.97$\pm$0.1  & 1.74$\pm$0.2 & 2.8(18)\\
          & \OI145      &  70.77$\pm$0.33 & 2.67$\pm$0.1  & 1.79$\pm$0.1 & \\
          & \thCO(6--5) &  36.63$\pm$0.22 & 2.71$\pm$0.1  & 1.65$\pm$0.1 & \\
          & \CeiO(3--2) &  19.25$\pm$0.09 & 3.08$\pm$0.1  & 1.09$\pm$0.1 & \\
	  & H$_2$\,S(2)  &   & & 1.5$\times10^{-4}$ \\
	  & H$_2$\,S(3)  &   & & 1.0$\times10^{-4}$ \\
(-33,6) & \thCII      &  5.33$\pm$0.42 &   2.55$\pm$0.1 &  3.08$\pm$0.3 & 3.8(18)\\
        & \OI145      & 50.77$\pm$0.47 &   2.45$\pm$0.1 &  2.55$\pm$0.1 &  \\
        & \thCO(6--5) & 12.62$\pm$0.68 &   3.41$\pm$0.1 &  1.03$\pm$0.1 & \\
        & \CeiO(3--2) &  6.73$\pm$0.11 &   3.06$\pm$0.1 &  0.90$\pm$0.1 & \\
	  & H$_2$\,S(2)$^a$  &   & & 1.0$\times10^{-4}$ \\
	  & H$_2$\,S(3)$^a$  &   & & 2.0$\times10^{-5}$ \\
(-31,-31) & \thCII      &   4.34$\pm$0.29 & 4.02$\pm$0.1 & 2.28$\pm$0.2 & 3.4(18)\\
          & \OI145      &  48.21$\pm$0.47 & 3.80$\pm$0.1 & 3.02$\pm$0.1 & \\
          & \thCO(6--5) &  11.48$\pm$0.60 & 3.50$\pm$0.1 & 0.97$\pm$0.1 & \\
          & \CeiO(3--2) &  10.83$\pm$0.10 & 3.10$\pm$0.1 & 0.90$\pm$0.1 & \\
	  & H$_2$\,S(2)$^a$  &   & & 1.2$\times10^{-4}$ \\
	  & H$_2$\,S(3)$^a$  &   & & 3.5$\times10^{-5}$ \\
(32, -23) & \thCII       &   3.34$\pm$0.35 & 4.25$\pm$0.2 & 2.74$\pm$0.3 & 3.0(18) \\
          & \OI145       &  19.17$\pm$0.50 & 4.35$\pm$0.1 & 2.93$\pm$0.1 & \\
          & \CeiO(3--2)  &   4.77$\pm$0.09 & 3.20$\pm$0.1 & 1.09$\pm$0.1 & \\
	  & H$_2$\,S(2)$^a$  &   & & 3.0$\times10^{-5}$ \\
	  & H$_2$\,S(3)$^a$  &   & & 1.1$\times10^{-5}$ \\

(60,-45) & \thCII       &  1.95$\pm$0.24 &  3.91$\pm$0.2  & 2.20$\pm$0.3 & 1.3(18)\\
         & \OI145       & 24.27$\pm$0.67 &  3.58$\pm$0.1  & 2.42$\pm$0.1 & \\
         & \CeiO(3--2)  &  6.50$\pm$0.10 &  3.23$\pm$0.1  & 0.81$\pm$0.1 & \\
	 & H$_2$\,S(2)$^a$  &   & & 8.8$\times10^{-5}$ \\
	 & H$_2$\,S(3)$^a$  &   & & 1.4$\times10^{-5}$ \\
\hline
\hline
\end{tabular}

$^a$ H$_2$ data taken from \citet{larsson2017} and intensities expressed  in
units of erg\,sec$^{-1}$\,cm$^{-2}$\,sr$^{-1}$.
\end{table}

\section{The S\,1 PDR}

\begin{figure}[h]
\centering
\includegraphics[width=0.48\textwidth]{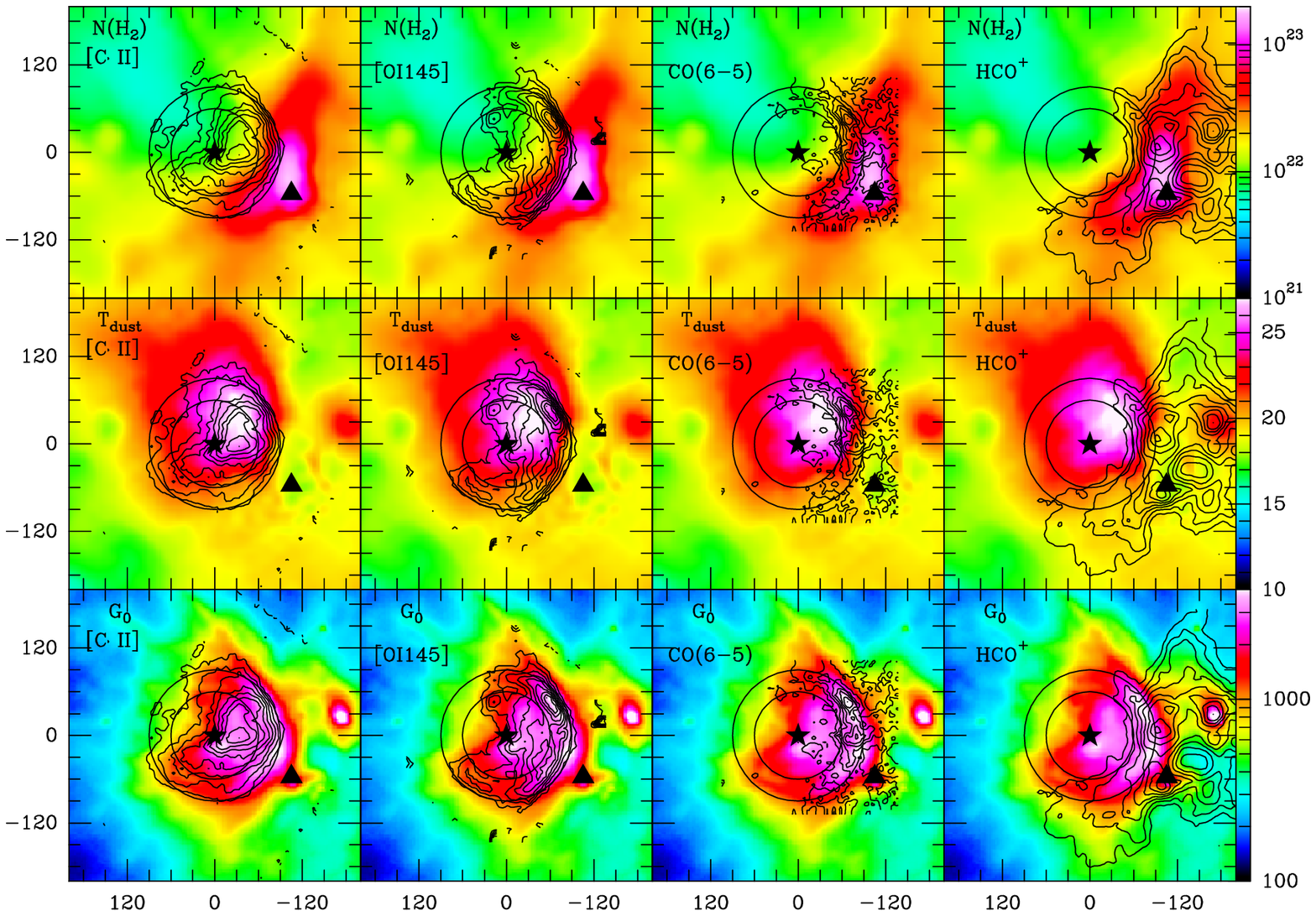}
\caption{Comparison of distribution of \CII, \OI\ 145\,\micron, CO(6--5)
and \hcoplus (4--3) emission vis-a-vis the column density, dust
temperature and FUV intensities estimated from PACS continuum data
observed as part of Herschel Gould Belt Survey. The color scales are
shown to the extreme right of each row of panels. The contour levels for
\CII, \OI\ 145\,\micron\ and \hcoplus (4--3) are at 20--100\% (in steps
of 10\%) of the peak values of 191, 88 and 14\,K\kms\ respectively. For
CO(6--5) the contours are at 40--100\% of the peak of 136\,\kms. {\rm
Circles drawn correspond to radii of 45\arcsec\ and 75\arcsec.}
\label{fig_overlay2}}
\end{figure}

Figure\,\ref{fig_overlay2} shows a comparison of the observed
distribution of the PDR and high density tracers with the molecular
hydrogen column density, dust temperature and FUV intensity, all derived
from dust continuum detected with PACS as part of the Herschel key
program on Gould Belt Survey \citep{Andre2010}. The column density and
dust temperature maps are directly taken from the Gould Belt Survey
website and we have estimated the FUV intensity from the observed
far-infrared (FIR) intensity as described in \citet{mookerjea2018}. The
circles drawn in Fig.\,\ref{fig_overlay2} are centered on S\,1 with
radii of 45 and 75\arcsec\ to guide the eye.  The column density peak in
the Oph\,A cloud is close to the position of VLA\,1623. The dust
temperature peaks at a position slightly offset from S\,1 and closer to
the \CII\ peak and  the embedded YSO LFAM\,9. The peak in the FIR
continuum map is located close to S\,1 and also to the south-west
\citep[Fig. A.1. in][]{mookerjea2018}. As indicated earlier, the PDR is
bound by the dense ambient cloud to the south-west and is more tenuous
to the north-east. Thus, the fraction of stellar FUV radiation
intercepted by the cloud is likely to be larger towards the south-west
than at positions towards the east and north-east which are radially
equidistant from S\,1.  The \CII\ traces the entire PDR gas, which is
also seen in the FUV map derived from the FIR continuum maps.  The \OI\
145\,\micron\ which has a  higher critical density preferentially picks
up the denser and warmer edge-on PDR rim to the west. The CO(6--5)
traces only the PDR clumps within a very narrow strip and the \hcoplus
(4--3) traces the higher density (and column density) molecular clouds
in the Oph\,A ridge, which also harbors the YSO VLA 1623.

Table\,\ref{tab_gauss} presents results of fitting Gaussian profiles to
the optically thin spectra of \thCII, \OI\ 145\,\micron, \CeiO (3--2) and
\thCO(6--5) at the positions already analyzed in Fig.\ref{fig_slab}. We
find that the \thCII\ and \OI145\ lines are significantly broader than the
\thCO(6--5) lines, except at the position (-61,45) which corresponds to
the peak of \OI\ 145\,\micron\ as well as \thCO(6--5). Additionally, the
central velocities of the PDR tracers are  red- and blue-shifted relative
to the molecular cloud tracer depending on whether the positions are to
the north or south of S\,1 respectively. The \CeiO(3--2) primarily traces
the ambient molecular cloud and hence typically peaks around 3.1\,\kms.

\subsection{The FUV field \label{sec_fuv}}

The distribution and emission from the PDR is primarily a function of
the FUV (6~eV $\leq$ h$\nu$ $<$ 13.6~eV) radiation field and volume
density of the PDR gas. The star S\,1 is the primary source of FUV
radiation for this PDR.  Based on the observed radio continuum flux
at 1420\,MHz we estimate the S\,1 to have a spectral type of B2.5--B3V.  We
thus used the Schmidt-Kaler relation for a B3V star ($T_{\rm
eff}$=18,700\,K and a radius of 4.15\,\rsun) and Kurucz model atmosphere
to estimate the FUV radiation field distribution considering only
projected distances and geometrical dilution.  The FUV field is
typically expressed in units of the \citet{Habing1968} value for the
average solar neighborhood FUV flux, 1.6$\times 10^{-3}$
ergs\,cm$^{-2}$\,s$^{-1}$.  We find that at a radial distances of
45\arcsec\ and 75\arcsec\ from S\,1 the unattenuated FUV field
from S\,1 is
2.70$\times 10^4$ and 8500\,G$_0$, respectively.  An alternative method
of estimating the strength of the FUV radiation in the region involves
the use of the observed total far-infrared (FIR) intensity, assuming
that the entire FUV energy is intercepted and absorbed by the grains and
is reradiated in the FIR.  We used the far-infrared observations of
Herschel/PACS to estimate the values of FUV radiation field
around S\,1
(Fig.\,\ref{fig_overlay2}). We find that the observed FIR distribution
is neither spherically symmetric around S\,1, nor does it peak at the
position of S\,1. The peak in FIR emission (coinciding with the peak
$T_{\rm dust}$) is primarily to the north-west, where the derived FUV
radiation is around 4000\,G$_0$ at a radius of 75\arcsec. To the east at
similar radii from S\,1 the estimated FUV emission is $\sim$
1200\,G$_0$.  By making a pixel to pixel comparison, we find that only for
the ridge-like structure to the north-west the FUV flux
predicted from the S\,1 FUV radiation and from the FIR continuum agree to
within a factor of 2. For regions between 40\arcsec\ to 75\arcsec\ from
S\,1 and at the ridge in the west, the two estimates differ by  up to a
factor of 10.  The discrepancy between the FUV radiation field derived
theoretically considering only geometric dilution and the field derived
indirectly from the observed far-infrared radiation can be due to: (a)
the emission in far-infrared continuum arising from regions, which are 
at much larger
distances than the projected distance used here (b) FUV radiation
escaping the region without being intercepted by material particularly
to the east and north-east, (c) presence of very high $A_{\rm V}$
clumps, which attenuate the FUV drastically but are too small to be
detected in single-beam continuum observations.

As discussed in Sec 3.1, the region is bound in the west and south-west
by the dense Rho Oph A ridge and possibly freely expanding to the
north-east. The structures visible in the far-infrared continuum images
primarily trace the column density of dust (and gas) along the lines of
sight, as is shown in the H$_2$ column density maps generated from the
same PACS maps (Fig.\,\ref{fig_overlay2}).  The lower levels of FIR
continuum emission closer to S\,1, as well as to the east is therefore a
result of lower column densities of dust (and molecular gas) in these
regions, while the higher FIR continuum emission to the north-west
indicate the presence of higher column density clumps, which is also
substantiated by the detection of \hcoplus (4--3) with JCMT and NH$_3$
with the Green Bank Telescope \citep{friesen2017}.

\subsection{Comparison of observed intensities with PDR models}

\begin{figure}[h]
\begin{center}
\includegraphics[width=0.50\textwidth]{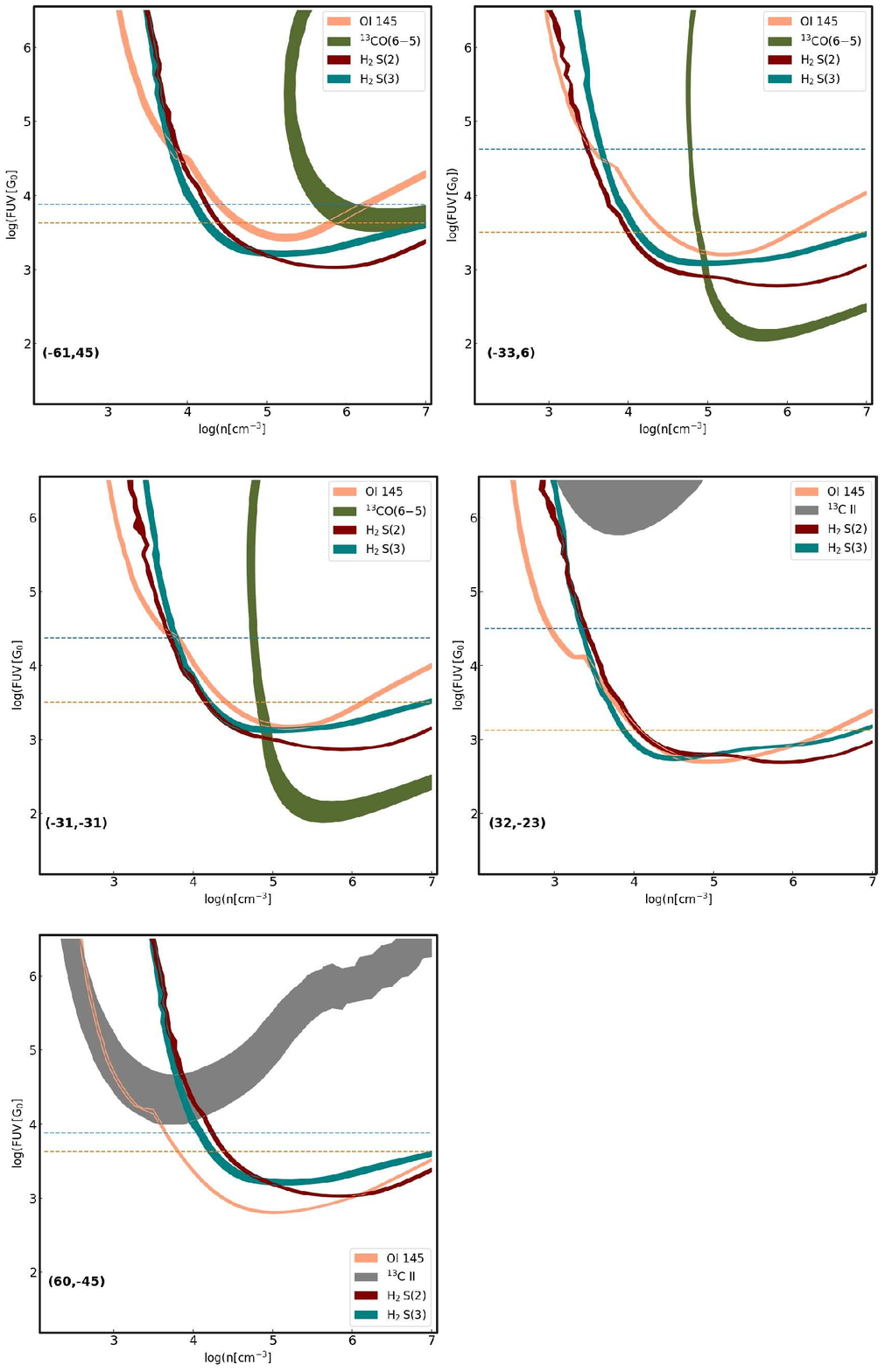}
\caption{Comparison of observations with prediction of line intensities as  function
of $n_{\rm H}$ and G$_0$ from an updated version of the \citet{Kaufman06}
PDR models (Wolfire private communication). For all positions except
(32,-23) and (60,-45) the \thCII\ intensities are beyond the values predicted by the
models. The horizontal dashed lines correspond to the values of G$_0$
estimated from the FIR intensities ($lower$) and from the stellar FUV
radiation scaled only by geometrical dilution due to the increase in
distance from S\,1.
\label{fig_wolfirepdr}}
\end{center}
\end{figure}

We compare the observed intensities of optically thin tracers \thCII, \OI\
145\,\micron\ and \thCO(6--5) with the predictions of plane-parallel
steady-state PDR models, which self-consistently calculate the intensities
as a function of the FUV flux and gas density of H nuclei $n_{\rm H}$.
These PDR models are from an updated version of the models by
\citet{Kaufman06} (Wolfire, private communication).  We perform the
analysis at the five selected positions, since these lie at different
radial distances from S\,1 to the east-west and north-south of S\,1, thus
tracing the distribution of the spectral lines arising from PDR reasonably
well. The S(2) and S(3) transitions of H$_2$ are produced by FUV
pumping, and hence the intensities are proportional to the FUV radiation. At the 
positions where the rotational lines of H$_2$ S(2) and S(3) have been
detected, we have also compared their intensities with the PDR model
predictions.  Figure\,\ref{fig_wolfirepdr} shows a comparison of the
observed intensities of the tracers (shown as contours) with the values
predicted by the models. The lower and upper limits of the FUV intensities
(shown by dashed horizontal lines in Fig.\,\ref{fig_wolfirepdr}) are
determined by the values derived from the FIR intensities and from the
stellar FUV radiation field, respectively (Sec.\,\ref{sec_fuv}). At most positions the observed
\thCII\ exceed the intensities predicted by the models for the entire
parameter space explored, hence the corresponding contours are not visible in
Fig.\,\ref{fig_wolfirepdr}. 

The critical densities (and E$_u$) for \thCII, \OI\ 145\,\micron\ and
CO(6--5) are 3000\,\cmcub\ (91\,K), 5.8$\times$10$^6$\,\cmcub\ (325\,K),
2.9$\times 10^5$\, \cmcub\ (116\,K), respectively.  The S(2) and S(3)
transitions of H$_2$ have critical densities (and E$_u$) of 2.2$\times
10^5$\,\cmcub\ (1682\,K) and 9.4$\times 10^5$\,\cmcub\ (2504\,K) ,
respectively.  Thus \OI145, \thCO(6--5) and S(2) H$_2$ transitions could
arise from PDR gas of similarly high densities, while \thCII\ traces the
low-density PDR. Though the H$_2$ lines have rather high critical densities
and high upper energy levels, these are produced by FUV pumping.  For
values of $G_0/n<10^{-2}$ the intensities of the S(2) and S(3) lines of
H$_2$ depend on the FUV intensities while for $G_0/n>10^{-2}$ the same
intensities are indicators of the densities \citep{Kaufman06}.  At the
selected positions, the FUV radiation field lies between
10$^3$--10$^{4.5}$, which implies that $G_0/n=10^{-2}$ corresponds to
$n$=10$^5$--10$^{6.5}$\,\cmcub.  For most positions the \OI145\ and H$_2$
line intensities indicate $n\sim10^4$\,\cmcub\ while \thCO(6--5) where
detected suggest $n>10^{5}$\,\cmcub. For most positions, the observed
\thCII\ intensities exceed the intensities predicted by the model
corresponding to the densities indicated by the \OI145\ and the estimated
range of FUV radiation fields, by factors of 1.2--3. The largest discrepancy,
a factor of 3--6, is seen at the position (-33,6). Additionally, the 
model predictions for the \OI145/\thCII\ intensity ratios indicate unrealistically
high (for \thCII\ emission) densities at all positions, that is not corroborated by
the high-density tracers such as CO(6--5) and HCO$^+$(4--3).

Comparison of the observed emission from the S\,1 PDR with models clearly shows the
contributions of gas at primarily three different density regimes,
10$^{5}$--$10^6$\,\cmcub, 10$^{3.5}$--10$^{4.5}$\,\cmcub\ and $<10^{3.5}$\,\cmcub,
although the \thCII\ emission is significantly under-produced by the models as is seen
both from the \thCII\ intensity and the \OI145/\thCII\ intensity ratio.
The highest density regions are traced by \thCO(6--5) and to some extent by \OI\
145\,\micron. Additionally, \OI\ 145\,\micron\ emission is excited in the medium
density gas as well, which is also traced by the H$_2$ S(2) and S(3) lines.  The most
diffuse component is primarily traced by the \thCII\  and for the positions (-33,6)
and (-61,45) the observed intensities, estimated to be only from the diffuse component
still far exceed the values predicted by the PDR models.

The lower \thCII\ intensities and correspondingly high \OI145/\CII\ ratio
predicted by the face-on uniform density plane-parallel PDR models can not
both be explained either by stacking layers of such PDRs along the
line-of-sight or by changing the viewing angle of the model. The most
plausible explanation for higher observed \thCII\ intensities relative to
\OI145\ line is in terms of the higher filling factor of the \thCII\
emission, which typically should have significant contribution from the
diffuse gas than from the high density gas emitting mostly in \OI145.  Such
discrepancies are also expected to arise from the shadowing effects of
clumps as well penetrability of non-uniform density PDRs consisting of
clump and inter-clump gas \citep{stutzki1988}. Use of three-dimensional PDR
models with inhomogeneity is needed, however such models also involve
additional parameters which need to be pre-determined using other
observational constraints. In the case of the S\,1 PDR, the self-absorption
of the main PDR tracers and the complex geometry of the region does not
allow us to further observationally constrain the parameters for such
clumpy PDR models. Qualitatively we can conclude that the fraction of
\thCII\ intensity putatively arising from diffuse PDR gas being higher
towards the west of S\,1 is likely to be an indicator of increased
clumpiness towards the west.

\section{Discussion \& Conclusion}

Presence of multiple emission and absorption components along the line of
sight towards the PDR associated with the S\,1 star in the $\rho$\, Ophiuchus
molecular cloud, results in complicated spectra that are difficult to
interpret. In this region, the emission arising from the PDR as well as
the molecular cloud overlap, most spectra are self-absorbed and there
are additional foreground filaments which criss-cross the region. The S\,1
PDR is restricted by the dense Oph\,A molecular cloud to the west and
south-west and appears to be expanding freely to the east.  The PDR is
tilted and somewhat warped with the front surface (facing the observer) of
the south-eastern side of the cavity being very dense and on the NW side
the cloud is denser on the back side. The gas distribution in the PDR is
rather inhomogeneous with clumps and ridges arising due to the disruption
of the dense ambient molecular cloud by the radiation from the star S\,1 and
also by the embedded YSOs. Analysis of the emission from the photon
dominated gas suggests the presence of at least three density components
consisting of high density ($10^6$\,\cmcub) clumps and  medium density
(10$^4$\,\cmcub) and diffuse ($10^{3}$\,\cmcub) interclump medium.

Using the velocity information and the optically thin spectra of \thCII\
and \OI\ 145\,\micron\ we have shown that the absorption features in
\CII\ and \OI\ 63\,\micron\ arise due to
the foreground layers of the same PDR. The ratio of column
densities of \cplus\ and O$^0$ in the diffuse foreground PDR layers, within
the limits of uncertainties introduced particularly by the assumed value of
the scaling factor for the \OI\ lines is between 2--3, which is comparable
to the solar [O]/[C] abundance ratio of 3.5. Spectral analysis of our data
allows to qualitatively constrain the temperature of the absorbing layers.
The presence of heavy foreground absorption in the \OI63\ spectra, but the
complete absence of self-absorption in the \OI145\ line profiles suggests a
low-excitation status of the foreground gas \-- low compared to the energy
of the OI $^3$P$_1$ level which is 227 K above ground.  On the other hand,
the prominent self-absorption in the CO(6--5) line points to a somewhat
elevated temperature of the absorbing gas, sufficient to populate the J=5
level (80\,K above ground).

The \CII\ spectra show the self-absorption dips even far to the east and
the rarer isotope $^{13}$\cplus\ has  been detected at a large number of
positions, suggesting an $N$(\cplus)$\sim$ of a few times
$10^{18}$\,\cmsq\ over an extended region. The estimated
$N$(\cplus) lie within the typical range of values between
10$^{18}$--10$^{19}$\,\cmsq\ that is found in Galactic PDRs
\citep{ossenkopf2013,mookerjea2019}. The H$_2$ column density estimated from dust
continuum emission maps range between 1--2.5\,10$^{22}$\,\cmsq\ at these
positions. This suggests values of [\cplus/H$_2$] ranging between
(1.5--4.8)\,10$^{-4}$ which leads to \cplus/H = (0.8--2.4)\,10$^{-4}$.
The derived value of \cplus/H is consistent with the value \cplus/H =
1.5$\times 10^{-4}$ obtained considering solar abundance with the
assumption that half of the carbon is in \cplus. The uncertainties in
the value of $N$(\cplus) arise due to the assumed values of \tkin\ used,
which were derived from the peak of the planck-corrected \CII\ spectra
(Sec. 3.4) and are likely a lower limit to the temperature. 

Based on the comparison of the observed intensities with the PDR models
(Fig.\,\ref{fig_wolfirepdr}) we have identified the range of densities
which could explain the \OI\ 145\,\micron\ intensities. We find that for
the \OI\ 145\,\micron\ peak, densities between
10$^{4}$--10$^{5}$\,\cmcub\ can explain the \OI145\ intensities. Using
non-LTE approximations, the observed intensities for these densities can
only be explained by \tkin$>100$\,K (consistent with our derived
\tkin$^{\rm min}$ of 157\,K) and for $N$(O$^0$) between (1.5--3)$\times
10^{19}$\,\cmsq. Similar considerations suggest that for the positions
(-33,6) and (-31,-31) the observed intensities can be explained by a
\tkin\ of 100\,K and $N$(O) of (2--3)$\times 10^{19}$\,\cmsq\ for
$n=$10$^5$\,\cmcub. For these two positions the
lower-density-higher-G$_0$ solution require \tkin = 120\,K and $N$(O) =
5$\times 10^{19}$\,\cmsq. However, we emphasize that it is likely  that
both the high- and low-density gas with possibly different filling
factors contribute to the emission. For positions to the west of S\,1,
as pointed out earlier, the FUV flux derived from FIR intensities is
lower than the value expected from the estimated stellar radiation since
part of the radiation is not intercepted by dust. For these positions
the higher FUV radiation values corresponding to the geometrically
diluted stellar radiation are likely to be a closer representation of
reality. The densities at these positions are thus $\sim 10^4$\,\cmcub,
which correspond to $N$(O)$\sim10^{19}$\,\cmsq\ for \tkin = 100\,K. Most
of the observed \OI\ 145\,\micron\ intensities can be explained by
$N$(O) between (1--3)$\times 10^{19}$\,\cmsq.  On the other hand, the
column density $N$(O) of the self-absorbing gas dominating the \OI\
63\,\micron\ profile is only one-tenth of the O$^0$ column density,
which shows up in emission.  The total $N$(O) estimated for the
S\,1 PDR is  similar to the column densities ($>10^{19}$\,\cmsq) seen in
dense molecular gas in sources like OMC-1 \citep{herrmann1997} and
L1689N \citep{caux1999}.  Typical values of $N$(O) estimated from
observations lie between 10$^{18}$--10$^{19}$\,\cmsq\
\citep{vastel2000,vastel2002}.

We estimate the gas pressure to be in the range of 10$^4$ --
10$^8$\,K\,\cmcub\ for densities between  10$^{3.5}$--
10$^6$\,\cmcub\ and temperatures, $T_{\rm kin}$, of 60--120 K
in the three gas components we identified.  The ambient high
density cloud that harbors the $\rho$\,Oph A region to the
west  with typical temperature of 10--20\,K and density of
10$^6$\,\cmcub\ \citep{Liseau2015} has a thermal pressure of
around 10$^7$\,\cmcub. Although we have detected
photoevaporation flows, no streaming motions indicative of
large pressure gradient were observed in the PDR and to the
west, where it interfaces with the high density molecular
cloud.  Interestingly, for an assumed temperature of  \tkin\
of 100\,K, the density of the PDR gas primarily contributing
to the thermal pressure and maintaining equilibrium at the
interface with the molecular gas, would be 10$^5$\,\cmcub.
This is consistent with the medium density interclump medium
as identified from our analysis of the emission from the PDR
gas.

\begin{acknowledgements}

The authors would like to thank W. Vacca for his help regarding the
stellar type of S\,1 and M. Wolfire for allowing the use of the updated
PDR models prior to publication.  BM acknowledges the support of the
Department of Atomic Energy, Government of India, under Project
Identification No. RTI 4002.  Based on observations made with the
NASA/DLR Stratospheric Observatory for Infrared Astronomy (SOFIA).
SOFIA is jointly operated by the Universities Space Research
Association, Inc.  (USRA), under NASA contract NAS2-97001, and the
Deutsches SOFIA Institut (DSI) under DLR contract 50 OK 0901 to the
University of Stuttgart. The development of GREAT was financed by the
participating institutes, by the Federal Ministry of Economics and
Technology via the German Space Agency (DLR) under Grants 50 OK 1102,
50 OK 1103 and 50 OK 1104 and within the Collaborative Research Centre
956, sub-projects D2 and D3, funded by the Deutsche
Forschungsgemeinschaft (DFG). This research has made use of the VizieR
catalog access tool, CDS, Strasbourg, France. The original description
of the VizieR service was published in A\&AS 143, 23.  This research
has made use of data from the Herschel Gould Belt survey (HGBS) project
(http://gouldbelt-herschel.cea.fr). The HGBS is a Herschel Key
Programme jointly carried out by SPIRE Specialist Astronomy Group 3
(SAG 3), scientists of several institutes in the PACS Consortium (CEA
Saclay, INAF-IFSI Rome and INAF-Arcetri, KU Leuven, MPIA Heidelberg),
and scientists of the Herschel Science Center (HSC).

\end{acknowledgements}

\end{document}